\documentclass[12pt]{article}

\usepackage[body={17.5cm, 21cm},right=2cm]{geometry}

\usepackage{color,graphicx}
\usepackage{epsf,setspace}
\usepackage{graphicx,epsfig}
\usepackage{amsmath, amssymb,amsfonts}
\usepackage{natbib}

\newcommand{\aap}{{\em Astron.\ Astrophys.}}
\newcommand{\apj}{{\em Astrophys.\ J.}}

\newcommand{\apjs}{{\em Astrophys.\ J.\ Suppl.}}
\newcommand{\jcap}{{\em J.\ Cosmol.\ Astropart.\ Phys.}}
\newcommand{\mnras}{{\em Mon.\ Not.\ R.\ Astron.\ Soc.}}
 
\newcommand{\physrep}{{\em Phys.\ Rep.}}
\newcommand{\prd}{{\em Phys.\ Rev.\ D}}
\newcommand{\prl}{{\em Phys.\ Rev.\ Lett.}}

\newcommand{\captionfonts}{\small}
\makeatletter  % Allow the use of @ in command names
\long\def\@makecaption#1#2{%
  \vskip\abovecaptionskip
  \sbox\@tempboxa{{\captionfonts #1: #2}}%
  \ifdim \wd\@tempboxa >\hsize
    {\captionfonts #1: #2\par}
  \else
    \hbox to\hsize{\hfil\box\@tempboxa\hfil}%
  \fi
  \vskip\belowcaptionskip}
\makeatother   % Cancel the effect of \makeatletter

\begin{document}

\begin{flushright} \mbox{\small HIP-2012-11/TH}
\end{flushright}
%\preprint{HIP-2012-11/TH}

\vspace{5mm}
\vspace{0.5cm}
\begin{center}

\def\thefootnote{\fnsymbol{footnote}}

{\Large \bf The Effect of Local non-Gaussianity \\ 
[0.5cm] 
on the Matter Bispectrum at Small Scales}
\\[2cm]
{\large  D.~G. Figueroa$^{a}$, E. Sefusatti$^{b,c}$,  A. Riotto$^{d}$ and F. Vernizzi$^{b}$}
\\[0.5cm]

\vspace{.3cm}
{\normalsize { \sl $^{a}$ Helsinki University and Helsinki Institute of Physics, PL\,64~~FIN-00014, Helsinki, Finland}}\\

\vspace{.3cm}
{\normalsize { \sl $^{b}$ Institut de Physique Th\'eorique, CEA, IPhT, 91191 Gif-sur-Yvette c\'edex, France,\\
and CNRS, URA-2306, F-91191 Gif-sur-Yvette c\'edex, France}}\\

\vspace{.3cm}
{\normalsize { \sl $^{c}$ The Abdus Salam International Centre for Theoretical Physics, Strada Costiera 11, I-34151 Trieste, Italy}}\\

\vspace{.3cm}
{\normalsize { \sl $^{d}$ Department of Theoretical Physics and Center for Astroparticle Physics (CAP)\\ 24 quai E. Ansermet, CH-1211 Geneva 4, Switzerland}}\\

%\vspace{.2cm}

\end{center}

\vspace{.8cm}

\hrule \vspace{0.3cm}
{\small  \noindent \textbf{Abstract} \\[0.3cm]
\noindent 
%We present the first estimate of the effect of local non-Gaussianity on the matter bispectrum at small scales in the Halo Model approach. 
%We compute the matter bispectrum at small nonlinear scales in the presence of primordial local non-Gaussianity using the halo model approach. In particular,  we include in the standard expression for the halo model bispectrum the non-Gaussian corrections to the halo profiles,  halo mass function and bias functions. We then compare our results to a large ensemble of numerical simulations with Gaussian initial conditions and with primordial non-Gaussianity for a local nonlinearity parameter $f_{\rm NL} =  100$, at redshift $z=0$ and $z=1$. Predictions for the galaxy bispectrum at small scales are also provided.
We compute the matter bispectrum in the presence of primordial local non-Gaussianity over a wide range of scales, including the very small nonlinear ones. We use the Halo Model approach, considering non-Gaussian corrections to the halo profiles, the halo mass function and the bias functions. We compare our results in the linear and mildly nonlinear scales to a large ensemble of Gaussian and non-Gaussian numerical simulations. We consider both squeezed and equilateral configurations, at redshift $z=0$ and $z=1$. For $z=0$, the deviations between the Halo Model and the simulations are smaller than 10\% in the squeezed limit, both in the Gaussian and non-Gaussian cases. The Halo Model allows to make predictions on scales much smaller than those reached by numerical simulations. For local non-Gaussian initial conditions with a parameter $f_{\rm NL} =  100$, we find an enhancement of the bispectrum in the squeezed configuration $k = k_3 = k_2 \gg k_1 \sim 0.01~h^{-1}$Mpc, of $\sim 15\%$ and $\sim 25\%$ on scales 
$k \sim 1~h^{-1}$Mpc, at $z = 0$ and $z = 1$ respectively. This is mainly due to the non-Gaussian corrections in the linear bias. Finally we provide a very simple expression valid for any scenario, {\em i.e.}~for any choice of the halo profile, mass and bias functions, which allow for a fast evaluation of the bispectrum on squeezed configurations.

\vspace{0.5cm}  \hrule
\def\thefootnote{\arabic{footnote}}
\setcounter{footnote}{0}

%\maketitle

%\date{\today}

\newpage 
\tableofcontents

% PRIMORDIAL POTENTIAL
\newcommand{\fix}{\Phi(\mathbf{x})}
\newcommand{\fiLx}{\Phi_{\rm L}(\mathbf{x})}
\newcommand{\fiNLx}{\Phi_{\rm NL}(\mathbf{x})}
\newcommand{\fik}{\Phi(\mathbf{k})}
\newcommand{\fiLk}{\Phi_{\rm L}(\mathbf{k})}
\newcommand{\fiLkone}{\Phi_{\rm L}(\mathbf{k_1})}
\newcommand{\fiLktwo}{\Phi_{\rm L}(\mathbf{k_2})}
\newcommand{\fiLkthree}{\Phi_{\rm L}(\mathbf{k_3})}
\newcommand{\fiLkfour}{\Phi_{\rm L}(\mathbf{k_4})}
\newcommand{\fiNLk}{\Phi_{\rm NL}(\mathbf{k})}
\newcommand{\fiNLkone}{\Phi_{\rm NL}(\mathbf{k_1})}
\newcommand{\fiNLktwo}{\Phi_{\rm NL}(\mathbf{k_2})}
\newcommand{\fiNLkthree}{\Phi_{\rm NL}(\mathbf{k_3})}

% CONVOLUTION RELATED QUANTITIES
\newcommand{\kernel}{f_{\rm NL} (\mathbf{k_1},\mathbf{k_2},\mathbf{k_3})}
\newcommand{\dirac}{\delta^{(3)}\,(\mathbf{k_1+k_2-k})}
\newcommand{\dirackonektwokthree}{\delta^{(3)}\,(\mathbf{k_1+k_2+k_3})}

% ABBREVIATIONS FOR THE MOST USED COMMANDS
\newcommand{\be}{\begin{equation}}
\newcommand{\ee}{\end{equation}}
\newcommand{\bea}{\begin{eqnarray}}
\newcommand{\eea}{\end{eqnarray}}

% ANGULAR PARTS
\newcommand{\angk}{\hat{k}}
\newcommand{\angn}{\hat{n}}

% TRANSFER FUNCTIONS
\newcommand{\tfnow}{\Delta_\ell(k,\eta_0)}
\newcommand{\tf}{\Delta_\ell(k)}
\newcommand{\tfone}{\Delta_{\el\ell_1}(k_1)}
\newcommand{\tftwo}{\Delta_{\el\ell_2}(k_2)}
\newcommand{\tfthree}{\Delta_{\el\ell_3}(k_3)}
\newcommand{\tffour}{\Delta_{\el\ell_1^\prime}(k)}
\newcommand{\deltatilde}{\widetilde{\Delta}_{\el\ell_3}(k_3)}

% MULTIPOLES
\newcommand{\alm}{a_{\ell m}}
\newcommand{\almL}{a_{\ell m}^{\rm L}}
\newcommand{\almNL}{a_{\ell m}^{\rm NL}}
\newcommand{\almone}{a_{\el\ell_1 m_1}}
\newcommand{\almLone}{a_{\el\ell_1 m_1}^{\rm L}}
\newcommand{\almNLone}{a_{\el\ell_1 m_1}^{\rm NL}}
\newcommand{\almtwo}{a_{\el\ell_2 m_2}}
\newcommand{\almLtwo}{a_{\el\ell_2 m_2}^{\rm L}}
\newcommand{\almNLtwo}{a_{\el\ell_2 m_2}^{\rm NL}}
\newcommand{\almthree}{a_{\el\ell_3 m_3}}
\newcommand{\almLthree}{a_{\el\ell_3 m_3}^{\rm L}}
\newcommand{\almNLthree}{a_{\el\ell_3 m_3}^{\rm NL}}

% SPHERICAL HARMONICS
\newcommand{\YLMstar}{Y_{L M}^*}
\newcommand{\Ylmstar}{Y_{\ell m}^*}
\newcommand{\Ylmstarone}{Y_{\el\ell_1 m_1}^*}
\newcommand{\Ylmstartwo}{Y_{\el\ell_2 m_2}^*}
\newcommand{\Ylmstarthree}{Y_{\el\ell_3 m_3}^*}
\newcommand{\Ylmstarfour}{Y_{\el\ell_1^\prime m_1^\prime}^*}
\newcommand{\Ylmstarfive}{Y_{\el\ell_2^\prime m_2^\prime}^*}
\newcommand{\Ylmstarsix}{Y_{\el\ell_3^\prime m_3^\prime}^*}

\newcommand{\YLM}{Y_{L M}}
\newcommand{\Ylm}{Y_{\ell m}}
\newcommand{\Ylmone}{Y_{\el\ell_1 m_1}}
\newcommand{\Ylmtwo}{Y_{\el\ell_2 m_2}}
\newcommand{\Ylmthree}{Y_{\el\ell_3 m_3}}
\newcommand{\Ylmfour}{Y_{\el\ell_1^\prime m_1^\prime}}
\newcommand{\Ylmfive}{Y_{\el\ell_2^\prime m_2^\prime}}
\newcommand{\Ylmsix}{Y_{\el\ell_3^\prime m_3^\prime}}

\newcommand{\comm}[1]{\textbf{\textcolor{rossos}{#1}}}
\newcommand{\lsim}{\,\raisebox{-.1ex}{$_{\textstyle <}\atop^{\textstyle\sim}$}\,}
\newcommand{\gsim}{\,\raisebox{-.3ex}{$_{\textstyle >}\atop^{\textstyle\sim}$}\,}

% BESSEL FUNCTIONS
\newcommand{\jl}{j_\ell(k r)}
\newcommand{\jlfourone}{j_{\el\ell_1^\prime}(k_1 r)}
\newcommand{\jlfivetwo}{j_{\el\ell_2^\prime}(k_2 r)}
\newcommand{\jlsixthree}{j_{\el\ell_3^\prime}(k_3 r)}
\newcommand{\jlsix}{j_{\el\ell_3^\prime}(k r)}
\newcommand{\jlthree}{j_{\el\ell_3}(k_3 r)}
\newcommand{\jlthreetau}{j_{\el\ell_3}(k r)}

% GAUNT INTEGRALS
\newcommand{\Gaunt}{\mathcal{G}_{\el\ell_1^\prime \, \el\ell_2^\prime \, 
\el\ell_3^\prime}^{m_1^\prime m_2^\prime m_3^\prime}}
\newcommand{\Gaunttwo}{\mathcal{G}_{\el\ell_1^\prime \, \el\ell_2^\prime \, 
\el\ell_3}^{m_1^\prime m_2^\prime m_3}}
\newcommand{\Gauntstardef}{\mathcal{\rm h}_{\el\ell_1 \, \el\ell_2 \, \el\ell_3}^{m_1 m_2 m_3}}
\newcommand{\Gauntstarone}{\mathcal{G}_{\el\ell_1 \, L \,\, \el\ell_1^\prime}
^{-m_1 M m_1^\prime}}
\newcommand{\Gauntstartwo}{\mathcal{G}_{\el\ell_2^\prime \, \el\ell_2 \, L}
^{-m_2^\prime m_2 M}}

\newcommand{\de}{{\rm d}}

% INTEGRATION VARIABLES
\newcommand{\dangn}{d \angn}
\newcommand{\dangk}{d \angk}
\newcommand{\dangkone}{d \angk_1}
\newcommand{\dangktwo}{d \angk_2}
\newcommand{\dangkthree}{d \angk_3}
\newcommand{\dk}{d^3 k}
\newcommand{\dkone}{d^3 k_1}
\newcommand{\dktwo}{d^3 k_2}
\newcommand{\dkthree}{d^3 k_3}
\newcommand{\dkfour}{d^3 k_4}
\newcommand{\dallk}{\dkone \dktwo \dk}

% FOURIER TRANSFORM
\newcommand{\FT}{ \int  \! \frac{d^3k}{(2\pi)^3} 
e^{i\mathbf{k} \cdot \angn \eta_0}}
\newcommand{\planewave}{e^{i\mathbf{k \cdot x}}}
\newcommand{\dallkfourier}{\frac{\dkone}{(2\pi)^3}\frac{\dktwo}{(2\pi)^3}
\frac{\dkthree}{(2\pi)^3}}

%BISPECTRA
\newcommand{\Bis}{B_{\el\ell_1 \el\ell_2 \el\ell_3}^{m_1 m_2 m_3}}
\newcommand{\Avbis}{B_{\el\ell_1 \el\ell_2 \el\ell_3}}
\newcommand{\fNL}{f_{\rm NL}}
\newcommand{\fNLl}{f_{\rm NL}^{\rm loc}}

%LINE OF SIGHT INTEGRATION
\newcommand{\los}{\mathcal{L}_{\el\ell_3 \el\ell_1 \el\ell_2}^{L \, 
\el\ell_1^\prime \el\ell_2^\prime}(r)}
\newcommand{\loszero}{\mathcal{L}_{\el\ell_3 \el\ell_1 \el\ell_2}^{0 \, 
\el\ell_1^\prime \el\ell_2^\prime}(r)}
\newcommand{\losone}{\mathcal{L}_{\el\ell_3 \el\ell_1 \el\ell_2}^{1 \, 
\el\ell_1^\prime \el\ell_2^\prime}(r)}
\newcommand{\lostwo}{\mathcal{L}_{\el\ell_3 \el\ell_1 \el\ell_2}^{2 \, 
\el\ell_1^\prime \el\ell_2^\prime}(r)}
\newcommand{\losfNL}{\mathcal{L}_{\el\ell_3 \el\ell_1 \el\ell_2}^{0 \, 
\el\ell_1 \el\ell_2}(r)}

%DANI's Commands (def --> Emiliano's)
\newcommand{\hI}{\hspace{1cm}}
\newcommand{\hVII}{\hspace{.7cm}}
\newcommand{\hV}{\hspace{.5cm}}
\newcommand{\Ms}{\, h^{-1} \, {\rm M}_{\odot}}
\def\kMpc{\, h \, {\rm Mpc}^{-1}}

%%%%%%%%%%%%%%%%%%%%%%%%%%%%%%%%%%%%%%%%%%%%%%%%%%%%%%
\def\d{d}
\def\O{{\mathcal O}}
\def\C{{\rm CDM}}
\def\me{m_e}
\def\te{T_e}
\def\ti{\tau_{\rm initial}}
\def\tci#1{n_e(#1) \sigma_T a(#1)}
\def\tr{\eta_r}
\def\dtr{\delta\eta_r}
\def\dd{\widetilde\Delta^{\rm Doppler}}
\def\dsw{\Delta^{\rm Sachs-Wolfe}}
\def\clsw{C_\ell^{\rm Sachs-Wolfe}}
\def\cldop{C_\ell^{\rm Doppler}}
\def\Dt{\widetilde{\Delta}}
\def\mut{\mu}
\def\vt{\widetilde v}
\def\hp{ {\bf \hat p}}
\def\sdv{S_{\delta v}}
\def\svv{S_{vv}}
\def\bvt{\widetilde{\bv}}
\def\delt{\widetilde{\delta_e}}
\def\cos{{\rm cos}}
\def\nn{\nonumber \\}
\def\bq{ {\bf q} }
\def\ba{ {\bf p} }
\def\bap{ {\bf p'} }
\def\bqp{ {\bf q'} }
\def\bp{ {\bf p} }
\def\bpp{ {\bf p'} }
\def\bk{ {\bf k} }
\def\bx{ {\bf x} }
\def\bv{ {\bf v} }
\def\qp{ p^{\mu}k_{\mu} }
\def\qpp { p^{\mu} k'_{\mu} }
\def\bgm{ {\bf \gamma} }
\def\bkp{ {\bf k'} }
\def\gq{ g(\bq)}
\def\gqp{ g(\bqp)}
\def\fp{ f(\bp)}
\def\h#1{ {\bf \hat #1}}
\def\fpp{ f(\bpp)}
\def\fz{f^{(0)}(p)}
\def\fpz{f^{(0)}(p')}
\def\f#1{f^{(#1)}(\bp)}
\def\fps#1{f^{(#1)}(\bpp)}
\def\dq{ {d^3\bq \over (2\pi)^32E(\bq)} }
\def\dqp{ {d^3\bqp \over (2\pi)^32E(\bqp)} }
\def\dpp{ {d^3\bpp \over (2\pi)^32E(\bpp)} }
\def\dtq{ {d^3\bq \over (2\pi)^3} }
\def\dtqp{ {d^3\bqp \over (2\pi)^3} }
\def\dtpp{ {d^3\bpp \over (2\pi)^3} }
\def\part#1;#2 {\partial#1 \over \partial#2}
\def\deriv#1;#2 {d#1 \over d#2}
\def\Done{\Delta^{(1)}}
\def\Dtwo{\widetilde\Delta^{(2)}}
\def\fone{f^{(1)}}
\def\ftwo{f^{(2)}}
\def\tg{T_\gamma}
\def\delpp{\delta(p-p')}
\def\delb{\delta_B}
\def\tc{\eta_0}
\def\DD{\langle|\Delta(k,\mu,\eta_0)|^2\rangle}
\def\DDL{\langle|\Delta(k=l/\tc,\mu)|^2\rangle}
\def\bkpp{{\bf k''}}
\def\kmkp{|\bk-\bkp|}
\def\kmkpsq{k^2+k'^2-2kk'x}
\def\tt{ \left({\tau' \over \tau_c}\right)}
\def\kt{ k\mu \tau_c}

\def\eq#1{Eq.\,(\ref{#1})}

\newcommand{\red}[1]{\textcolor{red}{#1}}
\newcommand{\white}[1]{\textcolor{white}{#1}}

%
%
%
%\renewcommand{\topfraction}{0.99}
%\renewcommand{\bottomfraction}{0.99}
%@twocolumnfalse\endcsname

%%%%%%%%%%%%%%%%%%%%%%%%%%%%%%%%%%%%%%%%%%%%%%%%%%%%%%%%%%%%%%%%%%%%%%%%%%%%
\section{Introduction}
\label{sec:intro}
%%%%%%%%%%%%%%%%%%%%%%%%%%%%%%%%%%%%%%%%%%%%%%%%%%%%%%%%%%%%%%%%%%%%%%%%%%%%

\noindent The possibility of detecting a non-Gaussian component of primordial origin in the Cosmic Microwave Background (CMB) temperature fluctations has been at the center of intense work during the past ten years, both in theoretical and in observational cosmology. Such detection would in fact provide crucial insight in the physics of the early Universe, possibly leading to rule-out the simplest model of canonical, single-field, slow-roll inflation \citep{BartoloEtal2004, KomatsuEtal2009A}. Current contraints from CMB observations are consistent with Gaussianity \citep{KomatsuEtal2011}. Still, CMB observations with the Planck mission \citep{PLANCK2006} will significantly improve such constraints, while approaching the limit for an ideal CMB experiment. At the same time, several ongoing and upcoming projects will map very large fractions of the matter and galaxy distributions by means of weak lensing and redshift measurements. This will offer the opportunity to probe and constrain, by a different mean, the 
non-Gaussianity of the initial conditions \citep{CarboneMenaVerde2010, GiannantonioEtal2011}.

In fact, for the {\em local model} of primordial non-Gaussianity \citep{SalopekBond1990}, constraints from the CMB and the Large-Scale Structure (LSS) from current datasets are already comparable \citep{SlosarEtal2008, XiaEtal2011}. This results from the rather unexpected effect that this specific model has on the bias of halos and galaxies. As shown by \citet{DalalEtal2008}, local non-Gaussianity determines a significant scale-dependent correction to the linear halo bias parameter, consequence of the strong correlation between the primordial large-scale and small-scale matter perturbations, the latter responsible for the collapse of dark matter halos. On the other hand, other models of non-Gaussianity or ``shapes'' of the initial curvature bispectrum provide corrections to a bias with a lesser or no scale-dependence, making the detection of this effect unfeasable by means of galaxy power spectrum measurements alone \citep{WagnerVerde2012, DesjacquesJeongSchmidt2011A, DesjacquesJeongSchmidt2011B, 
ScoccimarroEtal2012, SefusattiEtal2012}. 

However, the effects of generic non-Gaussian initial conditions on the matter and galaxy correlators, both in the two-point and higher-order correlation functions, are various and go beyond the halo bias correction mentioned above. In the first place, a non-vanishing, primordial curvature bispectrum is linearly evolved into a non-Gaussian component of the matter bispectrum, increasingly relevant at high-redshift and large scales when compared to the bispectrum induced by gravitational instability \citep[see {\em e.g.}][and references therein]{LiguoriEtal2010}. Its overall signal in future redshift surveys is large enough to provide constraints on non-Gaussian parameters competitive with those from the CMB \citep{ScoccimarroSefusattiZaldarriaga2004, SefusattiKomatsu2007}. Indeed, the number of Fourier modes available in large-volume three-dimensional redshift surveys is comparable to those corresponding to bidimensional CMB observations. 
 
In addition, primordial non-Gaussianity affects the {\em nonlinear} evolution of structure, leading to small-scales corrections to all matter correlators. The effect on the matter power spectrum has been explored computing one-loop corrections in standard cosmological Perturbation Theory (PT) in \citet{Scoccimarro2000A} for $\chi^2$ initial conditions and more recently in \citet{TaruyaKoyamaMatsubara2008} for the local and equilateral models. It has also been studied using more recent resummation schemes in PT by \citet{BernardeauCrocceSefusatti2010} and \citet{BartoloEtal2010}. Comparisons with numerical simulations can be found in \citet{PillepichPorcianiHahn2010} and \citet{SmithDesjacquesMarian2011}. The resulting corrections are small, of the order of the percent or below for the allowed values of the non-Gaussian parameters. The case of the matter bispectrum has been studied in standard PT by \citet{Sefusatti2009} and \citet{SefusattiCrocceDesjacques2010}. Here the correction to nonlinear clustering 
adds to the linear primordial component. For a local model with $\fNL=100$, it amounts to a few percent for generic triangles while reaching $\sim$ 20\% for squeezed configurations at redshift $z=1$. This is true also in the mildly nonlinear regime and, with small variations, at lower or higher redshift. However, PT techniques fail to describe the small scale regime, where matter perturbations become highly nonlinear. 

The Halo Model (HM) represents an approach to compute matter correlators in the nonlinear regime, beyond the reach of PT techniques. Besides, the HM easily allows for the inclusion of primordial non-Gaussian initial conditions. Indeed, the effects of non-Gaussianity on the matter power spectrum have been studied in the Halo Model framework in \citet{FedeliMoscardini2010} and in \citet{SmithDesjacquesMarian2011}, where measurements from N-body simulations were compared with predictions. They find that, for a local model with $\fNL=\pm 100$, the power spectrum is modified by about 3\% at $z=1$ and $k\sim 1\kMpc$, scale where the correction reaches a maximum.

In this work we provide the first computation of the effect of primordial local non-Gaussianity on the matter {\em bispectrum} in the highly nonlinear regime. In particular, we focus on squeezed triangular configurations, where one of the wavenumbers is much smaller than the other two. Such configurations are relevant, for instance, in constraining non-Gaussianity from weak lensing observations. Furthermore, for the local model of non-Gaussianity we naturally expect the squeezed limit to be mostly affected at large and intermediate scales. As shown in \citet{SefusattiCrocceDesjacques2010}, the effect of a local primordial non-Gaussianity on large scales is well described by the initial component of the matter bispectrum linearly extrapolated at the redshift of interest. Besides, in the mildly nonlinear regime, a local non-Gaussian component has a noticeable effect on the nonlinear evolution, captured in part by one-loop corrections in PT, an effect that is larger for squeezed configurations. We thus compare 
our HM predictions at linear and mildly nonlinear scales with measurements of the matter bispectrum in the simulations presented in \citet{SefusattiCrocceDesjacques2010}, and find a good agreement. Moreover, the HM approach adopted here allow us to reach highly non-linear scales. We also show that squeezed configurations can be captured by a significant simplification of the Halo Model expressions. We then provide a very simple set of formulae which allow, for any choice of the halo profile, mass and bias functions, for a fast computation of the bispectrum in the squeezed configuration. In principle, these results can be extended to describe the galaxy bispectrum at small scales, provided a recipe for the proper Halo Occupation Distribution. However, this approach is not free of certain caveats and certainly not at a satisfactory level of accuracy, so we discuss it only preliminarily in an appendix.

The paper is divided as follows. We introduce the definition of primordial non-Gaussianity in Sec.~\ref{sec:PNG}. The ingredients of the  Halo Model, including its non-Gaussian corrections, are discussed in detail in Sec.~\ref{sec:Halo Model}. In sec.~\ref{sec:sl} we focus on squeezed triangular configurations
and in Sec.~\ref{sec:asl} we present the mentioned simplified set of formulae. Equilateral configurations are discussed in Sec.~\ref{sec:eq}. Finally, our conclusions are drawn in Sec.~\ref{sec:conclusions}. We leave the description of the galaxy correlators for the appendix~\ref{sec:gbisp}. In Appendix~\ref{sec:AppendixB} we explain some technical issues regarding our calculations.

%%%%%%%%%%%%%%%%%%%%%%%%%%%%%%%%%%%%%%%%%%%%%%%%%%%%%%%%%%%%%%%%%%%%%%%%%%%%
\section{Primordial non-Gaussianity}
\label{sec:PNG}
%%%%%%%%%%%%%%%%%%%%%%%%%%%%%%%%%%%%%%%%%%%%%%%%%%%%%%%%%%%%%%%%%%%%%%%%%%%%

\noindent Measurements of the CMB bispectrum have confirmed, so far, the Gaussianity of the initial perturbations to a high degree \citep{KomatsuEtal2011}. Thus, any departure from Gaussian initial conditions is  expected to be small, with a possible non-Gaussian component in the curvature fluctuations roughly of the order of $10^{-4}$ or less in units of the primordial gravitational potential $\Phi$. We assume that a small amount of non-Gaussianity can be generically described, at leading order, by a non-vanishing curvature bispectrum, $B_\Phi(k_1,k_2,k_3)$, defined by $\langle\Phi_{\bk_1} \Phi_{\bk_2} \Phi_{\bk_3}\rangle = (2\pi)^3B_\Phi(k_1,k_2,k_3)\delta_D(\bk_1+\bk_2+\bk_3)$. The specific dependence of such bispectrum on the shape of the triangular configuration $\{k_1,k_2,k_3\}$ depends on the details of the inflationary model\footnote{Note that a primordial Bispectrum can be generated not only from inflation, but also at preheating \citep{EnqvistEtal2004, ChambersRajantie2008, BondEtal2009} or from 
cosmic defects~\citep{ReganShellard2010, FigueroaCaldwellKamionkowski2010}.} and on the post-inflationary evolution of curvature perturbations. 

In recent years, a large amount of work has been devoted to characterize the predictions of distinct models of inflation for the shape-dependence (or ``shape'', {\em tout court}) of the curvature bispectrum \citep[see {\em e.g.}][for recent reviews]{Chen2010, ByrnesChoi2010}. In this work we will focus on the local model \citep{SalopekBond1990}, which is defined in terms of a quadratic correction to the curvature perturbations, local in position space, given by \citep{KomatsuSpergel2001}
\be
\label{fnlphiin}
\Phi(\bx)=\phi(\bx)+\fNL\left[\phi^{2}(\bx)-\langle \phi^{2}(\bx) \rangle\right]\,,
\ee
where $\phi(\bx)$ is a Gaussian random field and the nonlinear parameter $\fNL$ measures the departure from Gaussian initial conditions. To first order in $\fNL$, \eq{fnlphiin} gives
\bea
B_\Phi(k_1,k_2,k_2)=2\,\fNL\,P_\phi(k_1)\,P_\phi(k_2)+ \rm cyc. \,,
\eea
where $P_{\phi}(k)$ is the power spectrum of the Gaussian component $\phi$, {\em i.e.}~$\langle\phi_{\bk} \phi_{\bk'}\rangle = $ $(2\pi)^3P_\phi(k)$ $\delta_D(\bk+\bk')$. Due the particularly simple definition of \eq{fnlphiin}, this model of primordial non-Gaussianity is probably the most popular and best studied. For instance, it allows for a straightforward implementation of local non-Gaussian initial conditions in numerical simulations, performed in the last few years by several groups \citep{GrossiEtal2007, DalalEtal2008, PillepichPorcianiHahn2010, DesjacquesSeljakIliev2009, WagnerVerdeBoubeker2010, LoVerdeSmith2011, ScoccimarroEtal2012}. From a theoretical point of view, the local bispectrum well approximates the bispectrum predicted by models where a non-Gaussian component is induced by mechanisms acting on perturbations outside the horizon, such as the curvaton model \citep{EnqvistSloth2002,LythUngarelliWands2003} and multiple-field inflation \citep{BartoloMatarreseRiotto2002, BernardeauUzan2002, 
VernizziWands2006}.

Present CMB limits on the local $\fNL$ parameter are given by $-10<\fNL <74$ \citep{KomatsuEtal2011}. Since the typical size of primordial fluctuations is of the order of $\phi\sim 10^{-5}$, it is easy to see from \eq{fnlphiin} that $\fNL\sim 100$ corresponds to a non-Gaussian correction of the order of $\sim 10^{-3}$. This simple observation justifies a treatment of non-Gaussian effects at linear order in $\fNL$, which we will assume throughout this work. Since the expression of \eq{fnlphiin} provides contributions to the curvature trispectrum of the order $\O\left(\fNL^2\right)$, which we neglect, we likewise ignore any effect due to a non-vanishing initial trispectrum. The validity of such an approximation for $|\fNL|\lesssim 100$ has been confirmed specifically for the matter and halo bispectrum in N-body simulations with local non-Gaussian initial conditions \citep{SefusattiCrocceDesjacques2010, SefusattiCrocceDesjacques2011}.

The most direct effect of primordial non-Gaussianity on matter correlators is given by the linear evolution of the primordial bispectrum. The linear matter density contrast $\delta_\bk(z)$ is related to the (constant) curvature perturbation $\Phi_\bk$ during matter domination via the Poisson equation, which we express as $\delta_\bk(z)=M(k,z)\,\Phi_\bk$, introducing the function
\be
M(k,z)\equiv \frac{2}{3}\frac{D(z)}{\Omega_m H_0^2}\,T(k)\,k^2\,. 
\ee
Here $T(k)$ is the matter transfer function, $\Omega_m$ and $H_0$ are the matter density in critical units and the Hubble rate today, and $D(z)$ is the linear growth function. 

The matter bispectrum is defined by $\langle\delta_{\bk_1} \delta_{\bk_2} \delta_{\bk_3}\rangle = (2\pi)^3B(k_1,k_2,k_3)\delta_D(\bk_1+\bk_2+\bk_3)$. At large scales it can be approximated by its leading order ({\em tree-level}) expression in PT, given by two components as
\be
\label{sum_B}
B(k_1,k_2,k_3)= B_0(k_1,k_2,k_3)+B_{\rm G}(k_1,k_2,k_3)\,.
\ee
The initial component $B_0$ is given by
\be
B_0(k_1,k_2,k_3)=M(k_1)\,M(k_2)\,M(k_3)\,B_\Phi(k_1, k_2,k_3)\,,
\label{NG}
\ee
where, as in what follows, we suppress the implicit time-dependence. In addition, the gravity-induced contribution $B_{\rm G}$ is given by
\be
\label{BG}
B_{\rm G}(k_1,k_2,k_3) = 2\,F_{2}(\bk_1,\bk_2)\,P_L(k_1)P_L(k_2) + {\rm 2~ perm.} \,,
\ee
with $P_L(k)\equiv M^2(k)\,P_\Phi(k)$ the linear matter power spectrum and $F_2$  the usual kernel representing the second-order solution in PT given by
\be
F_2(\bk_i,\bk_j)=\frac{5}{7} + \frac12\left({k_i\over k_j}+{k_j\over k_i}\right)(\hat{\bf k}_i\cdot\hat{\bf k}_j) + {2\over7}(\hat{\bf k}_i\cdot\hat{\bf k}_j)^2\,.
\ee
We refer the reader to the review \citet{BernardeauEtal2002} for an introduction to cosmological  PT.

At smaller scales the perturbative expansion for the matter bispectrum becomes more complicated as at next-to-leading order several one-loop corrections have to be taken into account \citep{Scoccimarro1997, Sefusatti2009}. For these nonlinear corrections we cannot separate the effect of the initial conditions from the effect of nonlinear gravitational evolution as the sum of two terms as in the tree-level expression \eq{sum_B}. Nevertheless the perturbative description shows that additional nonlinear corrections appear when initial higher-order correlators, like the bispectrum $B_0$, do not vanish. 

As we mentioned in the introduction, the scope of this work is to investigate how non-Gaussian initial conditions affect the nonlinear evolution of the matter bispectrum. While PT provides predictions for the nonlinear matter bispectrum with one-loop corrections improving significantly over the leading order terms given by \eq{BG}, such predictions are limited to the mildly nonlinear regime \citep[typically $k<0.15\kMpc$ for equilateral configurations at $z=1$,][]{SefusattiCrocceDesjacques2010}. To go beyond it is necessary to resort to numerical simulations or to phenomenological approaches such as the Halo Model, which we will discuss in the next section.

%%%%%%%%%%%%%%%%%%%%%%%%%%%%%%%%%%%%%%%%%%%%%%%%%%%%%%%%%%%%%%%%%%%%%%%%%%%%
\section{The Halo Model}
\label{sec:Halo Model}
%%%%%%%%%%%%%%%%%%%%%%%%%%%%%%%%%%%%%%%%%%%%%%%%%%%%%%%%%%%%%%%%%%%%%%%%%%%%

In its simplest formulation, the Halo Model \citep[see][for a review]{CooraySheth2002} assumes that all matter in the Universe belongs to dark matter halos, identified by their mass. Therefore, two distinct particles will either belong to the same halo or to two different ones. This picture allows one to compute the two-point correlation function (or the power spectrum) of density perturbation as the sum of two contributions: the 2-halo term, mainly accounting for the spatial correlations of the distribution of different halos in the Universe, and the 1-halo term which depends instead on the spatial distribution of matter inside a single halo. 
Clearly, while the 2-halo term is expected to describe large-scale correlations, the 1-halo term provides predictions in the nonlinear regime. 

More concretely, the HM expression for the %nonlinear 
matter power spectrum is given by
\be
\label{eq:HaloModelPowerSpectrum}
P(k)=P_{2h}(k)+P_{1h}(k)\,,
\ee
with the 2- and 1-halo contributions given by \citep{ScherrerBertschinger1991, Seljak2000, PeacockSmith2000, MaFry2000, ScoccimarroEtal2001A}
\bea
\label{P2h}
P_{2h}(k,z) & = & \frac{1}{\bar{\rho}^2}\left[\prod_{i=1}^2\int d\,m_i\,n(m_i,z)\,{\hat\rho}(k,m,z)\,\right] P_h(k,m_1,m_2)\,,\\
\label{P1h}
P_{1h}(k,z) & = & \frac{1}{\bar{\rho}^2}\int d\,m\,n(m,z)\,{\hat\rho}^2(k,m,z)\,,
\eea
where $\bar{\rho}$ is the mean matter density of the Universe, $n(m)$ is the halo mass function with $n(m)dm$ the number density of halos of mass between $m$ and $m+d m$, and $\hat{\rho}(k,m,z)$ is the Fourier transform of the spatial density profile $\rho(r,m)$ of a halo of mass $m$,  
\be
\hat{\rho}(k,m) = 4\pi\! \!\int \!\!dr \, r^2 \rho(r,m)\,{\sin(kr) \over kr}\,,
\ee
normalized so that $\hat{\rho}(0,M)=m$. The 2-halo term depends as well on the halo power spectrum, $P_h(k,m_1,m_2)$, describing the correlation between the centers of halos of mass $m_1$ and $m_2$. As we expect halos to be tracers of the underlying matter distribution, we can assume a linear bias relation between the halo and the matter density constrasts, so that $\delta_h \approx b_1\delta$. Thus, at large scales, the halo power spectrum can be approximated as
\be\label{eq:Ph}
P_h(k,m_1,m_2) = b_1(m_1)\,b_1(m_2)P_L(k)\,,
\ee
where $b_1(m)$ represents the linear bias function for halos of mass $m$. Note that for Gaussian initial conditions, $b_1$ only depends on the mass $m$, as implicitely assumed in \eq{eq:Ph}. However, if primordial non-Gaussianity is considered, $b_1$ will also depend in general on the scale $k$. Either way, the 2-halo term can be rewritten as
\be
P_{2h}(k,z)  =  \frac{1}{\bar{\rho}^2}\left[\prod_{i=1}^2\int d\,m_i\,n(m_i,z)\,{\hat\rho}(k,m,z)\,b_1(m,z)\,\right] P_L(k)\,,
\label{2h}
\ee
with an additional dependence on $k$ in the $b_1$ function for non-Gaussian initial conditions.

This description can be easily extended to the matter bispectrum. In the case of a three-point function, we should account for the possibility that the three points belong to just one, two or three dark matter halos. This means that there are now three distinct contributions to the Halo Model expression for the matter bispectrum, that is
\be
\label{BTOT}
B(k_1,k_2,k_3)=B_{3h}(k_1,k_2,k_3)+B_{2h}(k_1,k_2,k_3)+B_{1h}(k_1,k_2,k_3)\,,
\ee
where
\bea
\label{B3h}
B_{3h}(k_1,k_2,k_3,z) & = & {1\over\bar\rho^3}\left[\prod_{i=1}^3\int \!\!d\,m_i\,n(m_i,z)\,\hat{\rho}(m_i,z,k_i)\right] B_h(k_1,m_1;k_2,m_2;k_3,m_3;z)\,,\\
\label{B2h}
B_{2h}(k_1,k_2,k_3,z) & = & {1\over\bar\rho^3}\int\!\!d\,m\,n(m,z)\,\hat{\rho}(m,z,k_1)\int\!\!d\,m'\,n(m',z)\,\hat{\rho}(m',z,k_2)\,\hat{\rho}(m',z,k_3)\nonumber\\
 & & \times\, P_{h}(k_1,m,m',z)+{\rm cyc.}\,,\\
\label{B1h}
B_{1h}(k_1,k_2,k_3,z) & = & {1\over\bar\rho^3}\int\!\!d\,m\,n(m,z)\,\hat{\rho}(k_1,m,z)\,\hat{\rho}(k_2,m,z)\,\hat{\rho}(k_3,m,z)\,.
\eea
In this case, while the 2-halo term depends on the halo power spectrum as in the previous case, the 3-halo term involves the halo bispectrum, $B_{h}(k_1,m_1;k_2,m_2;k_3,m_3;z)$. Assuming a local bias relation between halos and matter, $\delta_h(m)=f(\delta)$, expanded perturbatively as $\delta_h(m)=b_1(m)\delta+[b_2(m)/2]\delta^2 + \mathcal{O}(\delta^3)$, it is possible to derive the tree-level expression for the halo bispectrum, valid only in the large-scale limit, 
in terms of the matter power spectrum $P(k)$ and bispectrum $B(k_1,k_2,k_3)$. This reads \citep{FryGaztanaga1993}
\bea\label{eq:Bh}
B_h(k_1,m_1;k_2,m_2;k_3,m_3;z) & = & b_1(m_1)\,b_1(m_2)\,b_1(m_3)\,B(k_1,k_2,k_3)\nonumber\\
& & + \left[ b_1(m_1)\,b_1(m_2)\,b_2(m_3)\,P(k_1)\,P(k_2)+{\rm cyc.}  \right] \,,
\eea
where $b_2(m)$ is the quadratic bias function. For Gaussian initial conditions $b_1$ and $b_2$ are scale independent. Moreover, since this equation is valid on large scales,  we can replace the matter power spectrum $P$ by its linear prediction $P_L$ and the matter bispectrum $B$ by its gravitational contribution $B_{\rm G}$, \eq{BG}, assuming a vanishing initial bispectrum $B_0=0$. If the initial conditions are {\em not} Gaussian the matter bispectrum $B$ also contains a non-vanishing initial contribution $B_0$, \eq{NG}, and the expressions of the halo power spectrum \eq{eq:Ph} and of the halo bispectrum \eq{eq:Bh} are modified, particularly for {\em local non-Gaussianity}, to account for scale-dependent corrections to both the linear and quadratic bias functions. 

Halo Model predictions for the real and redshift space matter bispectrum with Gaussian initial conditions have been recently studied and compared to numerical simulations by \citet{SmithShethScoccimarro2008} and \citet{ValageasNishimichi2011B}. \citet{SmithShethScoccimarro2008} presents a more detailed treatment of the dependence of halo model contributions on the large-scale PT expressions for the matter correlators which involve an explicit smoothing of the matter density by some filtering function. Here, for simplicity, we will ignore this issue as it does not affect the target of our calculations, that is the small-scale corrections induced by primordial non-Gaussianity. The model proposed by \citet{ValageasNishimichi2011B} differs from our as it is based on a Lagrangian description of halo correlations. Such description presents the advantage of predicting a 1-halo contribution that vanishes in the large-scale limit, as opposed to the constant asymptotic behavior resulting from \eq{B1h}, which has no 
physical justification. 
In fact, this large-scale 1-halo contribution is also present in the matter power spectrum and it is a well recognized problem of the halo model, for which no clear solution has been proposed so far \citep{CooraySheth2002, SmithEtal2003, CrocceScoccimarro2008}. Since here we are mainly interested in small scales effects we do not address it in our expressions. Finally, another issue related to the halo model description is given by halo-exclusion, that is the explicit accounting of the finite size of dark matter halos in halo correlation. For instance, a halo-exclusion correction has been implemented by \citet{SmithDesjacquesMarian2011} for the matter power spectrum, resulting in a few percent correction on intermediate scales ($k\sim 1\kMpc$). In that reference it is argued that the same correction could also approximately cancel the large-scale contribution from the 1-halo term. We have attempted to implement a similar correction for the halo bispectrum, but the resulting expression does not allow for an 
efficient numerical evaluation and we do not include it in our predictions.

It should be noted that many of the theoretical improvements proposed in the literature to the basic halo model description of Eq.~(\ref{B3h})-(\ref{B1h}) ultimately fail to reduce the discrepancy of the predictions with simulations results below a 10\% level in the mildly nonlinear regime. We therefore limit ourselves to the simplest expressions  and we leave any refinement for future work. The final results of this paper will in fact consist in small scales effects of non-Gaussian initial conditions, whose description already present a significant degree of complexity. Aside for the large-scale approximations for the matter correlators, we can clearly identify three essential ingredients in the Halo Model prescription for nonlinear correlators: the halo mass function, the halo bias functions and the halo profile. In the rest of this section we will describe each of them individually paying specific attention to the corrections induced by non-Gaussian initial conditions.

%%%%%%%%%%%%%%%%%%%%%%%%%%%%%%%%%%%%%%%%%%%%%%%%%%%%%%%%%%%%%%%%%%%%%%%%%%%%
\subsection{Halo Profiles}
\label{ssec:profile}

\noindent In the Halo Model formulation assumed here, each halo is labeled by a single parameter: its mass. The spatial distribution of matter in a halo of mass $m$ is specified by the halo density profile $\rho(r,m)$, interpreted as an average over all halos of the same mass. 

We will consider the Navarro, Frenk \& White (NFW) form for the halo density profile \citep{NavarroFrenkWhite1997},
\be
\rho(r)={\rho_s \over (r/r_s)(1+r/r_s)^2}\,,
\label{rho}
\ee
which assumes a universal profile as a function of $r$. The parameters $r_s$ and $\rho_s$ can be expressed in terms of the virial mass of the halo $m$ and the concentration parameter $c$. In particular, the virial mass is given by $m \equiv (4\pi/3)\,R_v^3\,\Delta_v\,\bar\rho$, with $R_v$ the virial radius, %$\bar\rho$ is the average matter density in the Universe  and we 
defined as the radius of a sphere within which the mean density of the halo is $\Delta_v$ times that of the Universe. we take $\Delta_v = 200$. The  concentration parameter $c$ is defined as $c= R_v/r_s$ and is typically a function of $m$. Here for $c$ we assume the form of \citet{BullockEtal2001}.
Thus
\be
r_s = \left( \frac{3 m}{4 \pi c^3 \Delta_v \bar\rho} \right)^{1/3} \; , \qquad \rho_s = \frac{1}{3} \Delta_v \bar\rho c^3 \left[ \ln(1+c) - c/(1+c)  \right]^{-1}\;,
\ee
where the second relation has been obtained by integrating Eq.~\eqref{rho} over the volume and using the definition of $m$ above.

So far, the effects of primordial non-Gaussianity on the profiles of dark matter halos have not been extensively investigated. To the best of our knowledge, the only measurements of halo profiles in numerical simulations with non-Gaussian initial conditions are those presented by \citet{SmithDesjacquesMarian2011}. In the case of a local $\fNL =\pm100$, they have found that the ratio between the non-Gaussian to Gaussian halo profile in real space can be fitted by 
\be\label{eq:rhoNGa}
%R_\rho(r,m,\fNL)\equiv
\frac{\rho_{\rm NG}(m,r,z,\fNL)}{\rho_{\rm G}(m,r,z)} \propto \left[1+y(m)\log_{10}{r\over r_X(m)}\right]\, .
\ee
%The normalization of ${\rho}_{\rm NG}$ can be obtained by imposing $\hat{\rho}_{\rm NG}(m,k\rightarrow0,z,\fNL)) \rightarrow m$, where a hat denotes the Fourier transform. 
The mass-dependence of the functions $y(m)$ and $r_X(m)$ is obtained by fitting the ratio above  over the limited mass-range $\sim 10^{13} - 10^{15}\Ms$, and further imposing $y(m_{\rm min}) = r_X(m_{\rm min}) = 0$, where $m_{\rm min}$ is the minimal mass, typically of order $m_{\min} \sim (10^{4}-10^{6})\Ms$, adopted in the numerical integrations in the Halo Model terms\footnote{This artificial low-mass cut-off must be introduced to avoid the computationally challenging integration of pushing the lower limit down to zero in the halo term expressions. See Appendix \ref{sec:AppendixB} for details.}. Note, however, that this constraint has no real physical motivation. 

\begin{figure}[t]
\includegraphics[width=0.98\textwidth]{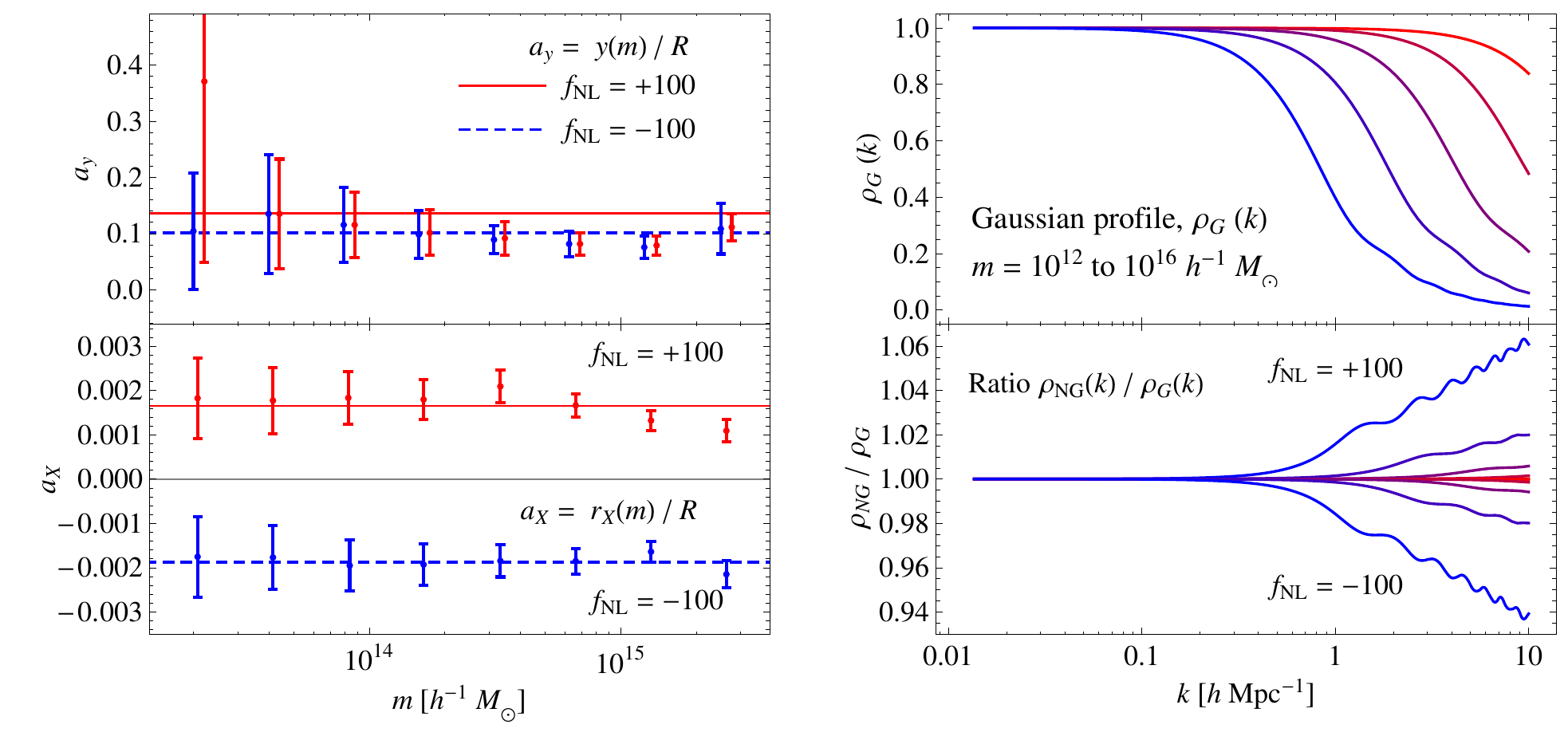}
\caption{{\em Left panel}: measurements of the parameter $a_y\equiv y(m)/R$ ({\em top}) and of the pivot point $a_X\equiv r_X(m)/R$ ({\em bottom}) describing the non-Gaussian correction to the halo profile, Eq.~(\ref{eq:rhoNGa}), derived from the plots of $y(m)$ and $r_X(m)$ shown in Fig.~8 of \citet{SmithDesjacquesMarian2011}, as a function of the mass $m$. Red and blue data points correspond respectively to $\fNL=+100$ and $-100$ with the red, continuous lines and blue, dashed lines being, respectively, the constant interpolating values. {\em Right panel}: Gaussian halo profile normalized to unity at large scales, $\hat\rho(k,m)/m$ ({\em top}), for $m = 10^{12}$, $10^{13}$, $10^{14}$, $10^{15}$ and $10^{16}\Ms$ ({\em right to left}) and the ratio between the non-Gaussian and Gaussian halo profiles ({\em bottom}) for the same masses (larger mass implies a larger correction) with $\fNL = \pm 100$. All curves are obtained for $z = 0$.
}
\label{fig:1}
\end{figure}
We have checked that both functions $y(m)$ and $r_X(m)$ approximately scale as $\propto R  = [3 m/(4\pi \bar\rho)]^{1/3}$, as one would expect. In particular, the data points in Fig.~8 of \citet{SmithDesjacquesMarian2011} can be fitted by the linear functions $y = {\rm sign}(\fNL)  a_y  R$ and $r_X = a_X  R$ with $a_y = 0.136$, $a_X = 0.00165$ for $\fNL = +100$ and $a_y = 0.110$, $a_X = - 0.00188$ for $\fNL = -100$. From this observation and assuming that the effect  of local non-Gaussianity  is proportional to $\fNL$, we decide to  parameterize the ratio \eqref{eq:rhoNGa}
by the following ansatz,
\be\label{eq:rhoNGb}
%R_\rho(r,m,\fNL)\equiv
\frac{\rho_{\rm NG}(m,r,z,\fNL)}{\rho_{\rm G}(m,r,z)} \propto \left[1-{\fNL\over 100}\,a_y R(m)\log_{10}{r\over a_X R(m)}\right]\,,
\ee
where the proportionality constant is choosen to ensure that the Fourier transform of the corrected profile verifies $\hat\rho_{\rm NG}(m,k\rightarrow0,z,\fNL) \rightarrow m$. Only more accurate simulations will be able to confirm or exclude this ansatz with more precision\footnote{While the NFW form for the Gaussian profile allows for an analytic Fourier transform, the non-Gaussian correction of \eq{eq:rhoNGa} or \eq{eq:rhoNGb} does not have a closed expression in Fourier space. We are then forced to obtain $\hat\rho(k,m,z,\fNL)$ numerically.}. The left panel of Figure~\ref{fig:1} shows the data points corresponding to the quantities $a_y$ and $a_X$ derived from Fig.~8 of \citet{SmithDesjacquesMarian2011}, plotted as a function of the halo mass and compared to our constant, best-fit values. For such quantities, no dependence on mass is required by the data. On the right panel of Figure~\ref{fig:1} we show the ratios of non-Gaussian to Gaussian halo profiles in Fourier space. They correspond to redshift $z = 
0$ and 
masses ranging from $m = 10^{12}$ to $10^{16}\Ms$, and were computed for $f_{\rm NL} = \pm 100$. As expected, the transition from $\hat\rho(m,k)/m = 1$ to $\hat\rho(m,k)/m \propto 1/(kr_s)^3$, around $k \sim 1/r_s$, is not modified in the presence of non-Gaussianity. Comparing the profiles with $f_{\rm NL} = -100, 0, 100$, for the same mass, we see that their relative difference is only noticeable in the large-momentum tails. The shift in amplitude is more appreciable for large masses. 

We can anticipate that the effect on the matter bispectrum due to the non-Gaussian corrections to the halo profiles is quite subdominant, particularly when compared to the effect of non-Gaussian corrections to the other ingredients of the HM, which we discuss in the rest of this section.

%%%%%%%%%%%%%%%%%%%%%%%%%%%%%%%%%%%%%%%%%%%%%%%%%%%%%%%%%%%%%%%%%%%%%%%%%%%%
\subsection{Halo Mass Function}
\label{ssec:MF}

The halo mass function $n(m)$ characterizes the number density of halos per unit mass. Theoretical predictions for this quantity go back to the early ansatz of \citet{PressSchechter1974} and have been largely improved over the years, either by comparing with N-body simulations \citep[{\em e.g.}~][]{JenkinsEtal1998, WarrenEtal2006, CrocceEtal2010, CourtinEtal011} and by more rigorous theoretical modeling in the framework of the excursion set theory \citep{BondEtal1991, ShethTormen1999, MaggioreRiotto2010A, MaggioreRiotto2010B, CorasanitiAchitouv2011A, CorasanitiAchituov2011B, ParanjapeLamSheth2012}.

The fraction of the total mass of the Universe contained in all the halos with mass in the range $[m, m+dm]$ can be written as
\begin{equation}
{1 \over \bar\rho}\, n(m)\, m\, dm = f(\nu)\,d\nu\,.
\end{equation}
The function $f(\nu)$ presents an approximately universal form and depends on the variable
\begin{equation}
\label{nu}
\nu\equiv \frac{\delta_c}{\sigma(m)}\,,
\end{equation}
with $\delta_{\rm c}$ representing the critical density for spherical collapses (we assume the Einstein-de Sitter value $\delta_c=1.68$) while $\sigma(M)$ is the r.m.s.~of matter fluctuations in spheres of radius $R = (3m/4\pi\bar\rho)^{1/3}$ (associated to the Fourier transform $W_R(k)$ of the top-hat function in real space), 
\be
\sigma^2(m)\equiv 4\pi\! \int\!dk \,k^2\,P_L(k)\,W_{R}(k)\,.
\ee
Since we will not be able to compare our results with numerical measurements on very small scales, we will omit implementing the most recent theoretical prescriptions for the mass function. Instead,  we will adopt  the Sheth \& Tormen~\citep[ST,][]{ShethTormen1999} expression, {\em i.e.}
\be
f(\nu) = A\sqrt{\frac{a\,\nu^2}{2\pi}}\left[1+\frac1{(a\,\nu^2)^{-p}}\right]e^{-a\nu^2/2}\,,
\ee
where $a=0.707$ and $p=0.3$ while $A = 0.322$ ensures a proper normalization.

The expression above will be modified in the case of non-Gaussian initial conditions. In particular, since primordial non-Gaussianity affects the high-mass tail of the halo mass function, such effect has been early recognized as a possible test of the initial conditions \citep[see][for a brief introduction to the early literature]{SefusattiEtal2007}. Moreover, thanks to several new results from numerical simulations assuming local non-Gaussian initial conditions \citep{DalalEtal2008, GrossiEtal2009, PillepichPorcianiHahn2010, DesjacquesSeljakIliev2009, WagnerVerdeBoubeker2010}, this topic has been recently subject of an intense theoretical activity \citep{LoVerdeEtal2008, AfshordiTolley2008, LamSheth2009B, Valageas2010, MaggioreRiotto2010C, DeSimoneMaggioreRiotto2011, DAmicoEtal2011, ParanjapeGordonHotchkiss2011, AchitouvCorasaniti2012, MussoParanjape2012}.

We  parameterize the effect of primordial non-Gaussianity in terms of the ratio between the $n_{\rm G}$ and $n_{\rm NG}$, respectively the mass function for Gaussian and non-Gaussian initial conditions, 
\be\label{eq:RNG}
R_{\rm NG}(m,z,\fNL) \equiv\frac{n_{\rm NG}(m,z,\fNL)}{n_{\rm G}(m,z)}\,.
\ee
Moreover, we assume for such function the expression proposed by \citet{LoVerdeEtal2008}, based on an Edgeworth expansion about the Press-Schechter mass function. At linear order in $\fNL$, this is given by
\be
R_{\rm NG}(\nu,\fNL)=1 + \frac16\,\nu\,(\nu^2 - 3)\, s_3(\nu) - \frac{1}{6}\, \left(\nu - \frac1\nu\right)\,\frac{ds_3(\nu)}{d\ln\nu}+{\mathcal O}(\fNL^2)\,,
\ee
where $s_3(\nu)$ represents the reduced skewness of the initial, smoothed, matter density field $\delta_R(\bx)$, defined as
\be
s_3(\nu)\equiv\frac{\langle\delta_R^3\rangle}{\langle\delta_R^2\rangle^{3/2}}=\frac{\langle\delta_R^3\rangle}{\sigma^3(m)}\,,
\ee
with the halo mass $m$ and radius $R$ related to the variable $\nu$ by \eq{nu}. The third order moment $\langle\delta_R^3\rangle$ is computed\footnote{In the evaluation of the mass function correction we make use of the following fit, 
\be
s_3(\nu)=\fNL\,\exp\left[-7.98 - 0.177 \ln\nu + 0.0186 \ln^2\nu - 0.00260\ln^3\nu\right]\,,
\ee
noticing that the dependence on scales -- {\em i.e.}~on the variable $\nu$ -- is relatively weak over the relevant range.} as an integral over the initial bispectrum, \eq{NG}. In addition, to improve the agreement between the measurements of the mass function correction in the numerical simulations of \citet{DesjacquesSeljakIliev2009} with the prediction of \eq{eq:RNG}, we introduce a scaling parameter $q$ defined by $R_{\rm NG}(\nu)\rightarrow R_{\rm NG}(q\nu)$ with $q=0.91$ \citep{SefusattiCrocceDesjacques2011}.

%%%%%%%%%%%%%%%%%%%%%%%%%%%%%%%%%%%%%%%%%%%%%%%%%%%%%%%%%%%%%%%%%%%%%%%%%%%%
\subsection{Halo Bias}
\label{ssec:bias}

The Halo Model expressions for the matter power spectrum and bispectrum involve, at large scales, a description of halos correlators, related in turn to the linear matter power spectrum and the tree-level matter bispectrum by an expansion of the halo bias relation. 

In the context of Gaussian initial conditions, the peak-background split approach \citep{ColeKaiser1989, MoWhite1996} provides a relation between the local halo number density and the large-scale matter perturbations, which can be turned into a dependence of the halo density contrast, $\delta_h$, on the local value of the matter density contrast, $\delta$, smoothed on a scale larger than the typical size of the halos considered. In this case, the relation $\delta_h(\delta)$ can be expanded in a Taylor series as \citep{FryGaztanaga1993}
\be\label{eq:biasexp}
\delta_h(\delta)=b_1(m)\,\delta+{1\over2!}\,b_2(m)\,\delta^2+\dots\,,
\ee
which defines the linear $b_1(m)$ and nonlinear $b_{i\geq 2}(m)$ halo bias functions. The bias functions can be derived from the unconditional halo mass function and, in the case of the ST form, one obtains for the first two the expressions \citep{ShethTormen1999, ScoccimarroEtal2001A}
\bea
\label{eq:b1G}
b_1(\nu) & = & 1+{a\,\nu^2 -1 \over \delta_{\rm c}}+{2p \over \delta_{\rm c}(1+(a\,\nu^2)^p)}\,,\\
\label{eq:b2G}
b_2(\nu) & = & \frac{8}{21}[b_1(\nu)-1]+\frac{a\,\nu^2}{\delta_c}\frac{a\,\nu^2-3}{\delta_c}+
\left(\frac{1+2p}{\delta_c}+2\frac{a\,\nu^2-1}{\delta_c}\right)\frac{2p/\delta_c}{1+(a\,\nu^2)^p}\,,
\eea
where $\nu=\nu(m)$ is defined by \eq{nu} while $a=0.707$ and $p=0.3$ are the ST values of two mass function parameters. 
The requirement for the total matter density to be given by
\bea
\rho(\bx) & \equiv & \bar{\rho}\left[1+\delta(\bx)\right] = \int \!dm\,m\,n(m)\left[1+\delta_h(m)\right]\nonumber\\
& = & \int \!dm\,m\,n(m)\left[1+\sum_i \frac{b_i(m)}{i!}\delta^i(m)\right] \,,
\eea
imposes the condition 
\be\label{eq:condrho}
\int\! dm\,m\,n(m)=\bar{\rho}\,,
\ee
along with the constraints on the bias functions,
\bea
{1\over\bar\rho}\!\!\int\!\! dm\,m\, n(m)\, b_1(m) &=& \int\!\! d\nu \,f(\nu)\, b_1(\nu) = 1\,,\\
{1\over\bar\rho}\!\!\int\!\! dm\,m\, n(m)\, b_i(m) &=& \int\!\! d\nu \,f(\nu)\, b_i(\nu) = 0, ~~~\forall\,\,i >1 \,,
\label{eq:condbias}
\eea
where $\delta_{ij}$ is the Kronecker delta. Such relations assure that, on large scales ($k \rightarrow 0$, $ \hat{\rho} \rightarrow m$), the 2-halo term of the power spectrum, \eq{P2h}, reduces to the linear power spectrum and the 3-halo term of the bispectrum, \eq{B3h}, reduces to the large-scale matter bispectrum.

The expansion \eqref{eq:biasexp} allows to derive perturbative bias expansions for halo correlation functions. Equations \eqref{eq:Ph} and \eqref{eq:Bh} represent the leading expressions for the halo power spectrum and bispectrum, respectively. 
These predictions  change significantly for non-Gaussian initial conditions. Indeed, since the bias functions are derivatives of the mass function, any correction to the latter induces a corresponding, scale-independent, correction to the bias. 

As first pointed-out by \citet{DalalEtal2008}, an initial local bispectrum is responsible as well for a {\em scale-dependent} correction to the linear bias. This rather unexpected effect received in the last few years a lot of attention since it allows to place tight constraints on the local $\fNL$ parameter from galaxy power spectrum measurements in current data-sets \citep[see, {\em e.g.}][]{SlosarEtal2008}. Several descriptions of the halo bias correction due to local non-Gaussianity have been proposed in the literature \citep[see the review in][and references therein]{DesjacquesSeljak2010B}. 
Here we will follow the approach of \citet{GiannantonioPorciani2010}, because it allows to easily derive an expression for the halo bispectrum \citep{BaldaufSeljakSenatore2011}, whose validity at large scales has been recently confirmed by comparison to numerical simulations in \citet{SefusattiCrocceDesjacques2011}. In this description, for local non-Gaussian initial conditions, the expansion \eqref{eq:biasexp} is replaced by a {\em bivariate} bias relation where the halo density contrast $\delta_h$ depends not only on the matter overdensity $\delta$, but also on the {\em Gaussian component} of the curvature perturbations $\phi$, defined by \eq{fnlphiin}. At linear order in $\fNL$ and up to second-order terms in $\delta$ and $\phi$, we have
\be\label{eq:biasexpNG}
\delta_h(\delta,\phi)=b_{10}\,\delta+b_{01}\,\phi+\frac12\,b_{20}\,\delta^2+\,b_{11}\,\delta\,\phi+\O(\fNL^2)\,,
\ee
where the bias parameters $b_{ij}$, with the indices $i$ and $j$ corresponding to the powers of $\delta$ and $\phi$, respectively, depend on the halo mass. 
From the bias relation above we can derive a rather lengthy expression for the halo bispectrum \citep{BaldaufSeljakSenatore2011,SefusattiCrocceDesjacques2011}. This expression can be considerably simplified by replacing  the Gaussian component of the curvature perturbation $\phi$ by its full non-Gaussian value $\Phi$. This is  well justified for the linear term $b_{01}\phi$, while it is an approximation for the term $b_{11}\delta \, \phi$  \citep[see][for a detailed discussion]{ScoccimarroEtal2012}. Note however that, as $b_{01}$ and $b_{11}$ are proportional to $\fNL$, this approximation is correct up to  $\O(\fNL^2)$ and thus consistent with our treatment. 

Thus, we consider the following bias relation in Fourier space
\be\label{eq:biasexpNGk}
\delta_h(m,k)=b_{1}(m,k)\,\delta_{\bk}+\frac12 \int d^3q_1\, d^3q_2\, \delta_D(\bk-\bq_{12})\,b_2(m,q_1,q_2)\,\delta_{\bq_1}\,\delta_{\bq_1}+\O(\fNL^2)\,,
\ee
where the non-Gaussian bias functions are now scale-dependent. In particular, we have
\be\label{e:b1}
b_1(m,k) = b_{1,{\rm G}}(m) + \Delta b_{1,{\rm NG}}^{(a)}(m)+\Delta b_{1,{\rm NG}}^{(b)}(m,k)+\O(\fNL^2)\,,
\ee
where $b_{1,{\rm G}}(m)$ corresponds to the Gaussian value given by \eq{eq:b1G}, while the second and third terms on the r.h.s.~represent, respectively, a scale-independent and a scale-dependent correction \citep{DesjacquesSeljakIliev2009,DalalEtal2008}, given by 
\begin{eqnarray}
\Delta b_{1,{\rm NG}}^{(a)}(m) & = & -\frac{\nu}{\delta_c}\frac{\partial }{\partial \nu}\ln R_{\rm NG}(\nu,\fNL)\,, 
\label{eq:db1NGa}\\
\Delta b_{1,{\rm NG}}^{(b)}(m,k) & = & \frac{2\,\fNL\,\delta_c\,\left(b_{1,{\rm G}}-1\right)}{M(k,z)}\,.
\label{eq:db1NGb}
\end{eqnarray}
Similarly, for the quadratic bias function $b_2(m,k_1,k_2)$ we have
\be\label{e:b2}
b_2(m,k_1,k_2) = b_{2,{\rm G}}(m) + \Delta b_{2,{\rm NG}}^{(a)}(m)+\Delta b_{2,{\rm NG}}^{(b)}(m,k_1,k_2)+\O(\fNL^2)\,,
\ee
where $b_{2,{\rm G}}(m)$ is the constant quadratic bias for Gaussian initial conditions, \eq{eq:b2G}, while the non-Gaussian correction is given again by a scale-independent and a scale-dependent contribution \citep{SefusattiCrocceDesjacques2011, GiannantonioPorciani2010},
\begin{eqnarray}
\Delta b_{2,{\rm NG}}^{(a)}(m) & = & \frac{\nu^2}{\delta_c^2}\frac{1}{R_{\rm NG}(\nu,\fNL)}\frac{\partial^2 R_{\rm NG}(\nu,\fNL)}{\partial \nu^2}+2\,\left(b_{1,{\rm G}}-1\right)\,\Delta b_{1,{\rm NG}}^{(a)}\,,\label{eq:db2NGa}\\
\Delta b_{2,{\rm NG}}^{(b)}(m,k_1,k_2) & = & 2\,\fNL\,\delta_{c}\left[\,b_{2,{\rm G}}+\left(\frac{13}{21}-\frac1{\delta_c}\right)\left(b_{1,{\rm G}}-1\right)\right]\left[\frac{1}{M(k_{1},z)}+\frac{1}{M(k_{2},z)}\right]\label{eq:db2NGb}\,.
\end{eqnarray}
Note that the conditions \eqref{eq:condrho}--\eqref{eq:condbias} are satisfied also when taking into account the corrections to the halo bias functions and the correction to the mass function, \eq{eq:RNG}. In particular, for $\fNL\ne 0$ we still have
\be\label{eq:condrhoNG}
\int \!\! dm\,m\,n_{\rm NG}(m,z,\fNL) = \bar{\rho}\,,
\ee
and
\be\label{eq:condbiasNG}
{1\over\bar{\rho}}\int \!\! dm\,m\,n_{\rm NG}(m,z,\fNL)\,b_i(m,z,\fNL,k) = \delta_{1i}\,.
\ee
As noted by \citet{SmithDesjacquesMarian2011}, the last relation implies
\be
\label{eq:condbiasNGsi}
{1\over\bar{\rho}}\int \!\! dm\,m\,n_{\rm NG}\,\left[b_{i,{\rm G}}+\Delta b_{i,{\rm NG}}^{(a)}\right]  =  \delta_{1i}\,,
\ee
involving only the scale-{\em independent} corrections, and also 
\be\label{eq:condbiasNGsd}
{1\over\bar{\rho}}\int \!\! dm\,m\,n_{\rm NG}\,\Delta b_{i,{\rm NG}}^{(b)} = 0 \,,
\ee
involving, instead, the scale-{\em dependent} part. At leading order in $\fNL$, \eq{eq:condbiasNGsd} is automatically satisfied  by the corrections in Eqs.~\eqref{eq:db1NGb} and  \eqref{eq:db2NGb}, which are proportional either to $[b_{1,{\rm G}}(m)-1]$ or to $b_{2,{\rm G}}(m)$\footnote{\citet{SmithDesjacquesMarian2011} point-out that, in general, the condition \eqref{eq:condbiasNGsd} is satisfied, for $i=1$, if $\Delta b_{1,{\rm NG}}^{(b)}(m,k) \propto (b_{1,{\rm G}}+\Delta b_{1,{\rm NG}}^{(a)}-1)$. Taking  into account this and the fact that at large $k$ the correction is negligible, they set $\Delta b_{1,{\rm NG}}^{(b)}=0$ in their calculation. While this approach is well justified, we keep the scale-dependent corrections to both linear and quadratic bias in our evaluation.}. We have now all the ingredients to compute the matter bispectrum for local non-Gaussianity.

%%%%%%%%%%%%%%%%%%%%%%%%%%%%%%%%%%%%%%%%%%%%%%%%%%%%%%%%%%%%%%%%%%%%%%%%%%%%
\section{The Matter Bispectrum}
\label{sec:mbisp}
%%%%%%%%%%%%%%%%%%%%%%%%%%%%%%%%%%%%%%%%%%%%%%%%%%%%%%%%%%%%%%%%%%%%%%%%%%%%

Let us now use  Eqs.~\eqref{BTOT}--\eqref{B1h} to compute the matter bispectrum. Including the corrections due to local non-Gaussianity, from \eq{eq:biasexpNGk} we can derive an expression for the non-Gaussian halo bispectrum in the tree-level approximation and plug it into \eq{B3h}. This is given by \eq{eq:Bh} with the bias functions $b_1(m)$ and $b_2(m)$ replaced by their non-Gaussian scale-dependent versions given in Eqs.~\eqref{e:b1} and \eqref{e:b2}, respectively. Furthermore, at the tree level the halo power spectrum to plug into \eq{B2h} is given by \eq{eq:Ph}, where for the linear bias we use \eq{e:b1}.

In the following, all mass integrals are computed numerically. Due to the divergent behavior of the integrands for $m \to 0$, they are computationally expensive. For this reason, we generalize a numerical prescription used to compute the matter power spectrum with the HM \citep{RefregierTeyssier2002, FedeliMoscardini2010}. We discuss this prescription in Appendix~\ref{sec:AppendixB}.

Since we compare our predictions with  measurements of the matter bispectrum in the numerical simulations presented in \citet{SefusattiCrocceDesjacques2010},
we  assume throughout the paper the same cosmology as these simulations: a flat $\Lambda$CDM model with matter and baryon content given by $\Omega_m=0.279$ and $\Omega_b=0.0462$, a spectral index $n_s=0.96$ and a r.m.s.~of fluctuations in spheres of $8$ {\em h}$^{-1}$ Mpc of $\sigma_8=0.81$. We  compute our predictions up to $k=10\kMpc$, at redshifts $z=0$ and $1$. However, the bispectrum measurements that we consider  are  limited to larger scales, {\em i.e.}  $k \lesssim 0.3 \kMpc$. 

%%%%%%%%%%%%%%%%%%%%%%%%%%%%%%%%%%%%%%%%%%%%%%%%%%%%%%%%%%%%%%%%%%%%%%%%%%%%
\subsection{The squeezed limit}
\label{sec:sl}

\begin{figure}[p]
\includegraphics[width=0.98\textwidth]{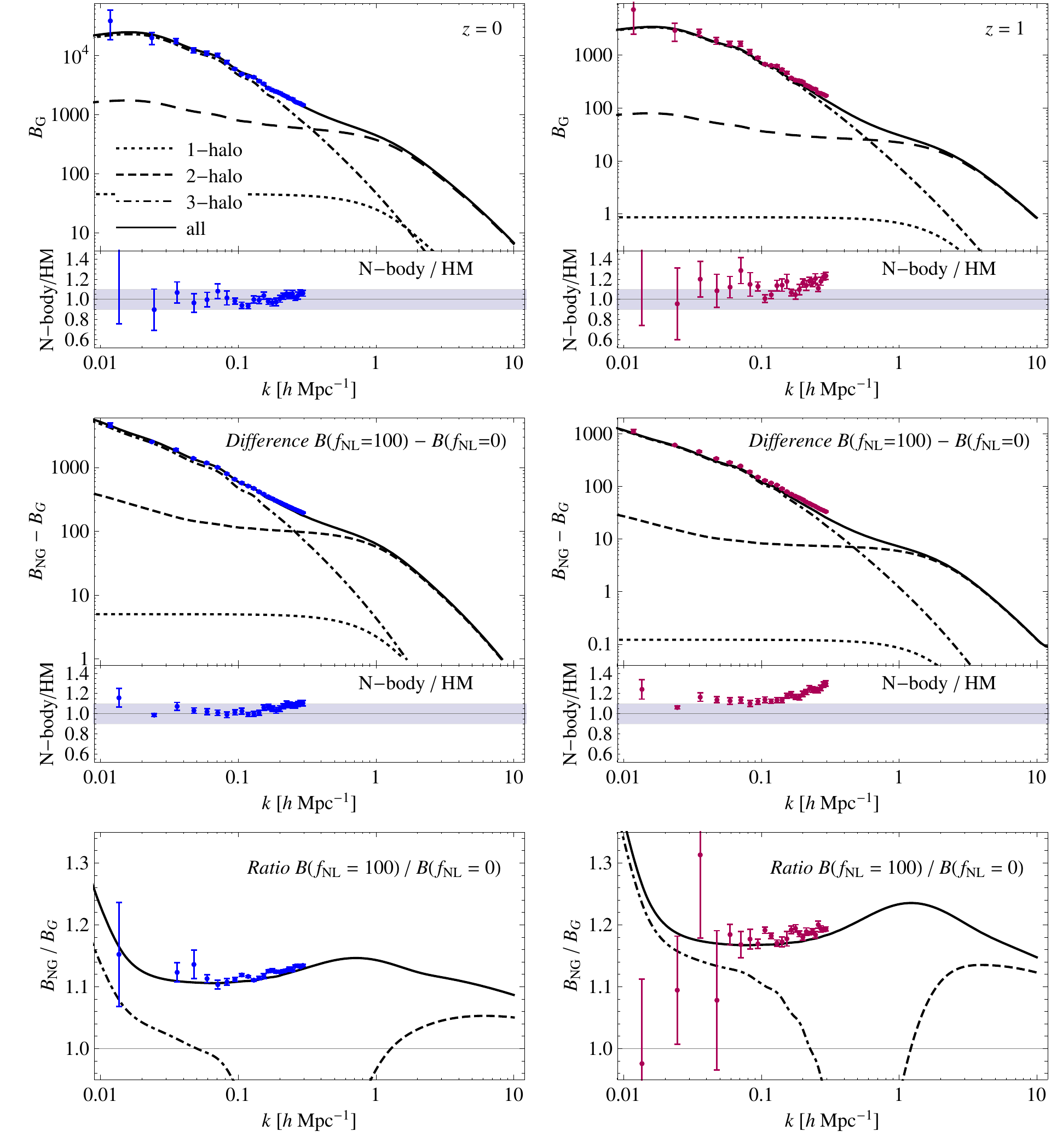}
\caption{
Squeezed configurations of the matter bispectrum, $B(k_1,k,k)$, with $k_1 \ll k_2 = k_3 \equiv k$, at redshift $z = 0$ ({\em left panels}) and $z=1$ ({\em right panels}). See text for explanation.}
\label{fig:BsmSq}
\end{figure}
Fig.~\ref{fig:BsmSq} shows the matter bispectrum for isosceles triangles, $B(k_1,k,k)$, with fixed $k_1=0.014 \kMpc$, as a function of $k$. Thus, this subset corresponds to increasingly squeezed configurations, starting with an equilateral one. Left panels show the results at redshift $z=0$, while right panels at $z=1$. Dotted, dashed and dot-dashed curves correspond, respectively, to the 1-halo, 2-halo and 3-halo contributions, while continuous lines correspond to their sum. In the first row we present the matter bispectrum for Gaussian initial conditions, together with the ratio between the measurements in N-body simulations and the model. The shaded area corresponds to a 10\% agreement. Our model is tested against simulations only in the mildly nonlinear regime, where we observe agreement within a 10\% error. The second and last rows show the correction due to non-Gaussian initial conditions. In particular,  the second row shows the {\em difference} between the predictions with and without non-Gaussianity,
 $\Delta B_{\rm NG}\equiv B_{\rm NG}(\fNL=100)-B_{\rm G}(\fNL=0)$, while the third row shows their {\em ratio},  $B_{\rm NG}(\fNL=100)/B_{\rm G}(\fNL=0)$. As shown in the middle row, the ratio between the N-body simulations and our predictions is inside the shaded area corresponding to a 10\% agreement for $z=0$ and slightly outside for $z=1$. This is a very good agreement, if we consider that the non-Gaussian correction for $\fNL=100$ is {\em per se} of the order of $\sim$ 10\% ($\sim$ 20\%) for $z = 0$  ($z = 1$). 
%Finally, in the lower plots we show  the {\em ratio} between the non-Gaussian and Gaussian predictions. 

At large $k$ the bispectrum is dominated by $B_{2h}(k_1,k,k)$. This is not surprising because this term is dominated by the position-space configuration in which two points are in the same halo and the third one is far away. We will take advantage of this fact in the following section.

Notice also that the relatively large corrections at intermediate scales due to both non-Gaussianity and nonlinear evolution, studied by \citet{SefusattiCrocceDesjacques2010}, extend  according to the Halo Model predictions well into the nonlinear regime, $k \ge 0.3\kMpc$. Similarly to what happens for the effect on the matter power spectrum \citep{FedeliMoscardini2010, SmithDesjacquesMarian2011}, the non-Gaussian to Gaussian ratio reaches a local maximum around $k \sim 1\kMpc$. In particular, such maximum represents a distorsion of the Bispectrum of the order $\sim 15\%$ at $k \approx 0.8 \kMpc$, for $z = 0$, and of the order of $\sim 25\%$ at $k \approx 1.4 \kMpc$, for $z = 1$. The enhancement in the matter bispectrum ($\sim 15\%-25\%$) due to the presence of local non-Gaussianity is therefore almost an order of magnitude greater than the analogous enhacement in the power spectrum ($\sim 2.5\%-3.5\%$).

In order to understand the role played by each of the non-Gaussian corrections affecting the Halo Model ingredients, in Fig.~\ref{fig:breakdown} we show such individual components both for the matter power spectrum ({\em left panel}) and for the squeezed configurations of the matter bispectrum ({\em right panel}). For the matter power spectrum the corrections to the halo mass function dominate the intermediate scales $k \sim 1\kMpc$, enhancing the power on these scales and inducing a bump in the ratio between the non-Gaussian and Gaussian power spectrum \citep{FedeliMoscardini2010, SmithDesjacquesMarian2011}, mentioned above. The corrections in the halo profile are only important on smaller scales, while those to the linear bias are always subdominat\footnote{Note that the effect $\propto 1/k^2$ due to non-Gaussian corrections in the linear bias does not blow up the amplitude of the matter power spectrum at large scales thanks to the constraint \eq{eq:condbiasNGsd}, which applies at those scales.}. For the 
matter bispectrum, the initial non-Gaussian conditions dominate its amplitude at large scales, as one can see on the right panels of Fig.~\ref{fig:breakdown}. However in this case, as opposed to the power spectrum, the corrections due to the linear bias function $b_1$ dominate the intermediate and small scales, {\em i.e.}~$k \gtrsim 0.1 \kMpc$. At $k \sim 1\kMpc$, also the corrections to halo mass function become important. Remarkably, over the range of scales considered here, the correction on the density profile is subdominant with respect to the corrections to the mass function or halo bias.

In conclusion, the enhancement %of power observed 
in the power spectrum due to the non-Gaussian corrections, is also present in the bispectrum. However, in the bispectrum %such a distortion 
it is dominated by the non-Gaussian corrections to the linear bias, and not by the corrections to the mass function as in the power spectrum. This translates into a bump in the ratio between the non-Gaussian and Gaussian bispectra, reaching a maximum amplitude in the squeezed configuration around $k \sim 1 \kMpc$ of $\approx 1.15$ for $z = 0$ and $\approx 1.25$ for $z = 1$.

\begin{figure}[t]
\includegraphics[width=0.98\textwidth]{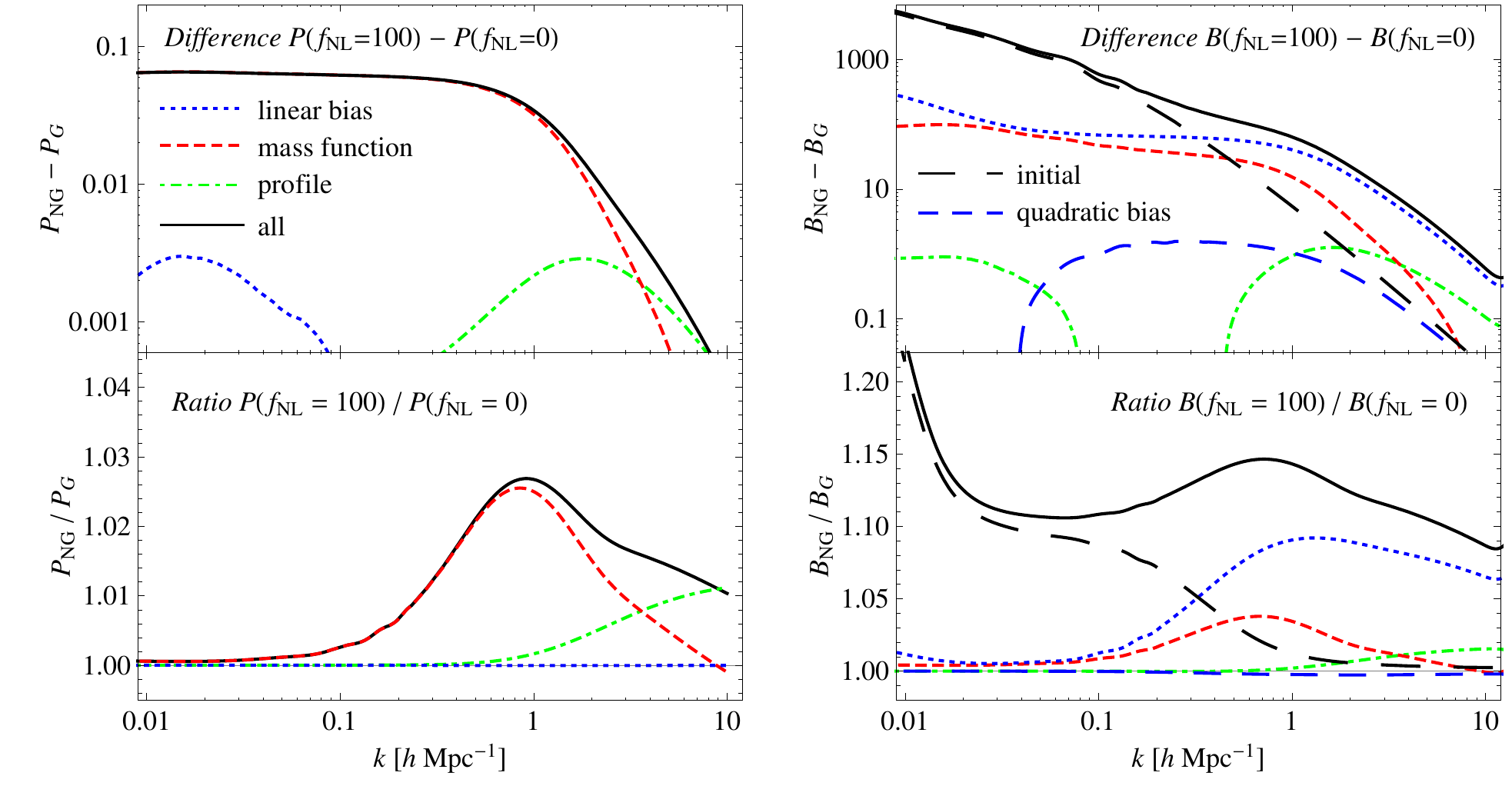}
\caption{Non-Gaussian corrections to individual ingredients of the Halo Model for the matter power spectrum ({\em left panel}) and for the squeezed configurations of the matter bispectrum ({\em right panel}). In addition to the full Halo Model ({\em continuous curve}) we consider the results of allowing for a non-Gaussian correction only in each one of the ingredients of the Halo Model: the mass function ({\em short-dashed curve}), the halo density profile ({\em dot-dashed}), the linear halo bias ({\em dotted}) and, for the bispectrum in particular, the quadratic halo bias ({\em medium-dashed}) and the initial component to the tree-level bispectrum ({\em long-dashed}). The correction to the quadratic bias ({\em upper right panel}) is shown with the sign changed. All panels share the same label. }
\label{fig:breakdown}
\end{figure}

%%%%%%%%%%%%%%%%%%%%%%%%%%%%%%%%%%%%%%%%%%%%%%%%%%%%%%%%%%%%%%%%%%%%%%%%%%%%
\subsection{Approximating the squeezed limit}
\label{sec:asl}

The squeezed limit allows for  a significant simplification of the Halo Model expressions. As already mentioned, in this limit the largest contribution to the bispectrum comes from the position-space configuration where two points are close and belong to the same halo while the third one is at larger distance from the first two, and hence is likely to belong to another halo. In this case we  expect the Halo Model prediction to be dominated by the 2-halo contribution, with the 1-halo and 3-halo terms being subdominant. In Fig.~\ref{fig:BsmSq}, where $k_1=0.014\kMpc$, it can be clearly appreciated that $B_{2h}$ becomes dominant over the other two terms for $k \gtrsim 0.3\kMpc$. At $k \gtrsim 1\kMpc$ $(B_{1h}+B_{3h})$ contributes $\sim$ 10\% of the total bispectrum, while at $k\simeq 10\kMpc$ $B_{3h}$ contributes less than 1\% and $B_{1h}$ less than $5\%$. This happens both for the bispectrum with Gaussian initial conditions and for the non-Gaussian component, $\Delta B_{\rm NG}$. Thus, the following discussion 
will be valid in the squeezed limit, independently of the initial conditions.

In particular, let us consider the squeezed (not necessarily isosceles) configuration $k_1 \ll k_2 \simeq k_3 \equiv k$. If $k_1$ is still in the linear regime, then we can safely set that the Fourier transform of the halo profile is
\begin{equation}\label{eq:rho_kIR}
\hat{\rho}(m,z,k_1)\simeq m\,
\end{equation}
in all bispectrum terms involving $k_1$ in Eqs.~\eqref{B3h}, \eqref{B2h} and \eqref{B1h}. By making this substitution, the expressions for the 1-, 2- and 3-halo terms greatly simplify. Moreover, by using the conditions in \eq{eq:condrho} and \eq{eq:condbias}, together with \eq{eq:rho_kIR}, the Halo Model bispectrum contributions become, at leading order in  $k_1 / k \ll 1$, 
\begin{align}
\label{pp1}
B_{1h}(k_1,k,k)&= 
{1 \over \bar{\rho}}\,\epsilon_2^{[m]}(k,\fNL)\,,\\
\label{pp2}
B_{2h}(k_1,k,k)&= 
\epsilon_2^{[b_1]}(k,\fNL)\,P_L(k_1)\,,\\
\label{pp3}
B_{3h}(k_1,k,k)&= 
2\left[ \frac{13}{14} + \left( \frac{4}{7} - \frac12 \frac{d \ln P_{L}}{d \ln k}\right) (\hat \bk_1 \cdot \hat \bk)^2 + \frac{\epsilon_{1}^{[b_2]}(k,\fNL)}{\epsilon_{1}^{[b_1]}(k,\fNL)} + \frac{2\,\fNL}{M(k_1,z)}\right]\, P_L(k_1)\,P_{2h}(k)\,,
\end{align}
with $\frac{d \ln P_{L}}{d \ln k}$ evaluated at $k$. The functions $\epsilon_i^{[F]}$ in these expressions
are defined as
\be
\epsilon_i^{[F]}(k,\fNL) \equiv \frac{1}{\bar{\rho}^{\, i}}\int\!\! dm\,n_{\rm NG}(m,z,\fNL)\,\hat{\rho}^{\,i}(m,z,k,\fNL)\,F(m,z,\fNL)\,,
\ee
where $F(m,z,\fNL)$ represents a generic function of mass and redshift. Thus, these functions are like an ``average'' of the function $F$, weighted by the mass function and the $i$th power of the Fourier transform of the density profile.
The first two terms inside the bracket of Eq.~\eqref{pp3} have been derived by taking the squeezed limit $k_1 \ll k_2 \approx k_3$ of $B_{\rm G} (k_1,k_2,k_3)$ in Eq.~\eqref{BG}, {\em i.e.},
\begin{equation}\label{eq:GaussSqueezed}
\begin{split}
B_{\rm G} (k_1,k_2,k_3) & \simeq 2 \left[ F_2(\bk_1,\bk_2) P_L(k_1) P_L(k_2) + F_2(\bk_1,\bk_3) P_L(k_1) P_L(k_3) \right]  \\
&= 2\left[ \frac{13}{14} + \left( \frac{4}{7} - \frac12 \frac{d \ln P_{L}}{d \ln k_2}\right) (\hat \bk_1 \cdot \hat \bk_2)^2 + {\cal O}(k_1/k_2) \right] P_L(k_1) P_L(k_2),
\end{split}
\end{equation}
where we have expanded $k_3=|\bf{k}_1+\bf{k}_2|$ for small $k_1/k_2$. Note that only $B_{3h}$, Eq.~\eqref{pp3}, depends on the angle between the long and the short mode. For isosceles configurations $\hat \bk_1 \cdot \hat \bk_2 \sim {\cal O}(k_1/k_2)$ and the angular dependence drops.

%In the particular case of isosceles squeezed configurations $k_1 \ll k_2 = k_3 \equiv k$, the following relation holds exactly
%\begin{equation}\label{F2Isos}
%F_2(\bk_1,\bk_2) + F_2(\bk_1,\bk_3) = {13\over14} - {5\over14}\left({k_1\over k}\right)^2\,.
%\end{equation} 
%Since the second term on the $rhs$ of Eq.~(\ref{F2Isos}) is just of order $\mathcal{O}\left(k_1/k\right)^2 \ll 1$, %the term proportional to $(\hat \bk_1 \cdot \hat \bk_2)^2$ can be simply written as
%%\begin{eqnarray}
%%&& \left(\frac{4}{7} - \frac12 \frac{d \ln P_{L}}{d \ln k}\right)(\hat \bk_1 \cdot \hat \bk_2)^2 = - {5\over14}\left({k_1\over k}\right)^2
%%\end{eqnarray} 
%%Since $k_1 \ll k$, this can then be simply neglected and 
%then we can drop it and approximate $B_{3h}$ simply as
%\begin{equation}\label{pp3Isos}
% B_{3h}(k_1,k,k) = 
%2\left[ \frac{13}{14} + \frac{\epsilon_{1}^{[b_2]}(k,\fNL)}{\epsilon_{1}^{[b_1]}(k,\fNL)} + \frac{2\,\fNL}{M(k_1,z)}\right]\, P_L(k_1)\,P_{2h}(k)\,,
%\end{equation}
%valid only for the isosceles squeezed configurations.

Eqs.~\eqref{pp1}--\eqref{pp3} %/\eqref{pp3Isos} clearly hold 
also hold for Gaussian initial conditions. In this case the term $2\,\fNL/M(k_1,z)$ inside the bracket drops and $n_{\rm NG}(m,z,\fNL)$, $\hat{\rho}(m,z,k,\fNL)$ and $F(m,z,\fNL)$ are replaced by their corresponding Gaussian expressions. Moreover, these equations can be also extended to other types of non-Gaussian initial conditions, provided that the corrections to $n_{\rm NG}$, $\hat{\rho}$, $b_1$ and $b_2$ are known. The derivation is analogous to the one given here for local non-Gaussianities. %Moreover, these equations can be also employed for generic non-Gaussian initial conditions, {\em i.e.}~different from the local one, provided that expressions for $n_{\rm NG}(m,z,\fNL)$, $\hat{\rho}(m,z,k,\fNL)$ and $F(m,z,\fNL)$ are known%\footnote{The term ${4\fNL\over M(k_1,z)}P_L(k_1)\,P_{2h}(k)$ should be replaced by $M(k_1,z)B_\Phi(k_1,k,k)[P_{2h}(k)/P_\Phi(k)]$ for isosceles squeezed configurations, with $P_{\Phi}$ the power spectrum and $B_{\Phi}$ the bispectrum characterizing the non-Gaussian model considered. For non-isosceles squeezed configuration one should look more carefully at the corresponding expression.}.

\begin{figure}[t]
\includegraphics[width=0.98\textwidth]{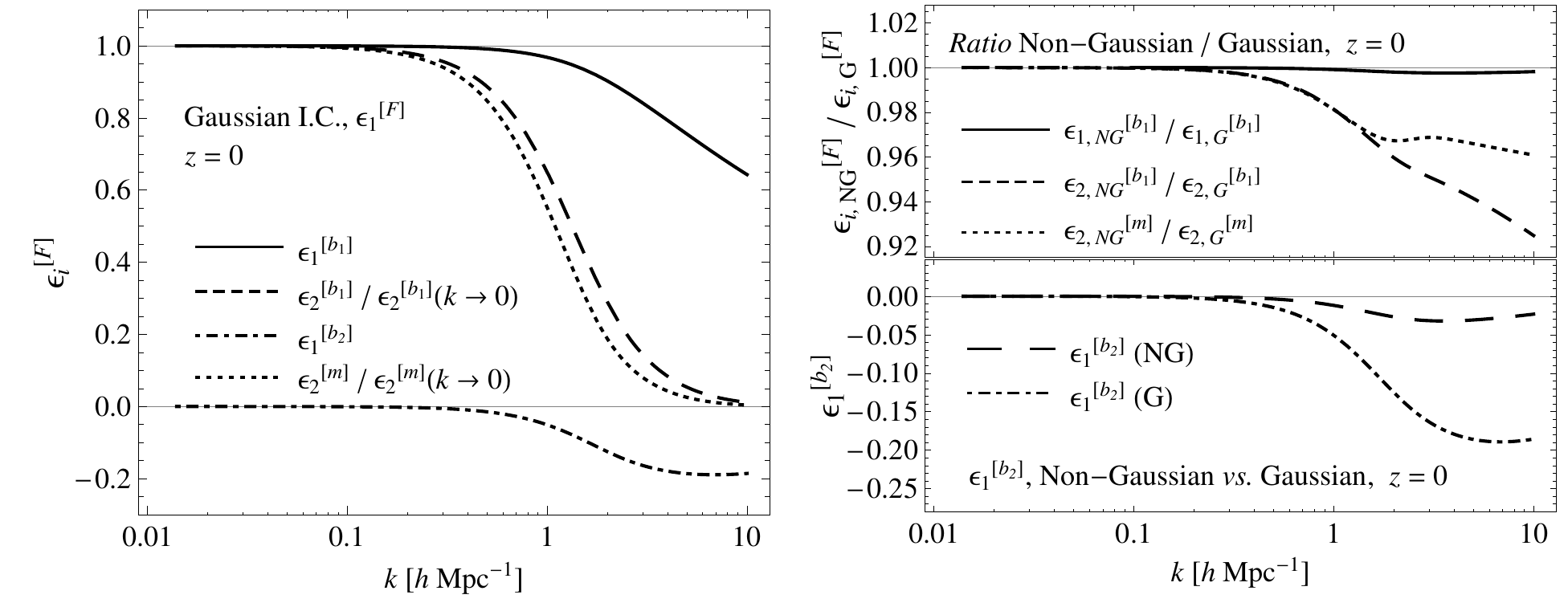}
\caption{{\em Left panel}: functions $\epsilon_1^{[b_1]}(k)$ ({\em continuous curve}), $\epsilon_2^{[b_1]}(k)/\epsilon_2^{[b_1]}(0)$ ({\em dashed curve}), $\epsilon_1^{[b_2]}(k)$ ({\em dot-dashed curve}) and $\epsilon_2^{[m]}(k)$ ({\em dotted curve}) as a function of $k$, evaluated for Gaussian initial conditions at redshift $z=0$. {\em Right, top panel}: non-Gaussian to Gaussian ratio for the functions $\epsilon_1^{[b_1]}(k)$ ({\em continuous curve}), $\epsilon_2^{[b_1]}(k)$ ({\em dashed curve}). {\em Right, bottom panel}: comparison between the function $\epsilon_2^{[b_1]}(k)/\epsilon_2^{[b_1]}(0)$ for Gaussian initial conditions ({\em dot-dashed curve}) and non-Gaussian initial conditions ({\em long dashed curve}).
}
\label{fig:eps}
\end{figure}
In Fig.~\ref{fig:eps} ({\em left panel}) we plot the functions $\epsilon_i^{[F]}$ involved in Eqs.~\eqref{pp1}--\eqref{pp3}, evaluated for Gaussian initial conditions. As for small  $k$, $\hat{\rho}(m,z,k) \rightarrow m$, then $\epsilon_1^{[b_1]}$ and $\epsilon_1^{[b_2]}$ reduce to the constraints of Eqs.~\eqref{eq:condrho}--\eqref{eq:condbias}, {\em i.e.}~$\epsilon_1^{[b_1]} \rightarrow 1$ and $\epsilon_1^{[b_2]} \rightarrow 0$. For $k \lesssim 0.2\kMpc$, $\epsilon^{[b_1]}_{1}$ only deviates less than $0.1\%$ from unity while $|\epsilon^{[b_2]}_{1}|$ is smaller than $0.001$. From these observations we conclude that the approximations above, Eqs.~\eqref{pp1}--\eqref{pp3}, are safely valid when $k_1 \lesssim 0.2\kMpc$. On the right panels of Fig.~\ref{fig:eps} we consider instead the effects of non-Gaussianity on the $\epsilon_i^{[F]}$ functions. In particular, the upper right panel shows the ratio between $\epsilon_i^{[b_1]}$ for $\fNL=100$ to the same quantities for Gaussian initial conditions. We notice 
that the effect on $\epsilon_1^{[b_1]}$ is very small, below one percent over the range of scales considered, while for $\epsilon_2^{[b_1]}$ it is more significant, of the order of 5\% at large $k$. The lower right panel compares $\epsilon_1^{[b_2]}$ for $\fNL=100$ and $\fNL=0$, illustrating that non-Gaussian initial conditions affect significantly this function at small scales.

In Fig.~\ref{fig:app} we show a comparison between the full Halo Model prediction ({\em thin, black curves}) and the approximate expressions presented above ({\em thick, blue curves}) for squeezed configurations with $k_1=0.014\kMpc$ as a function of $k_2=k_3=k$, at redshfit $z=0$. The left panels present the matter bispectrum with Gaussian initial conditions $B_{\rm G}$, while the right panels present the non-Gaussian component, $\Delta B_{\rm NG}\equiv B_{\rm NG}-B_{\rm G}$. The lower panels show the ratio between the approximate and the full expressions, distinguishing the 1-halo, 2-halo and 3-halo contributions. 
\begin{figure}[t]
\includegraphics[width=0.98\textwidth]{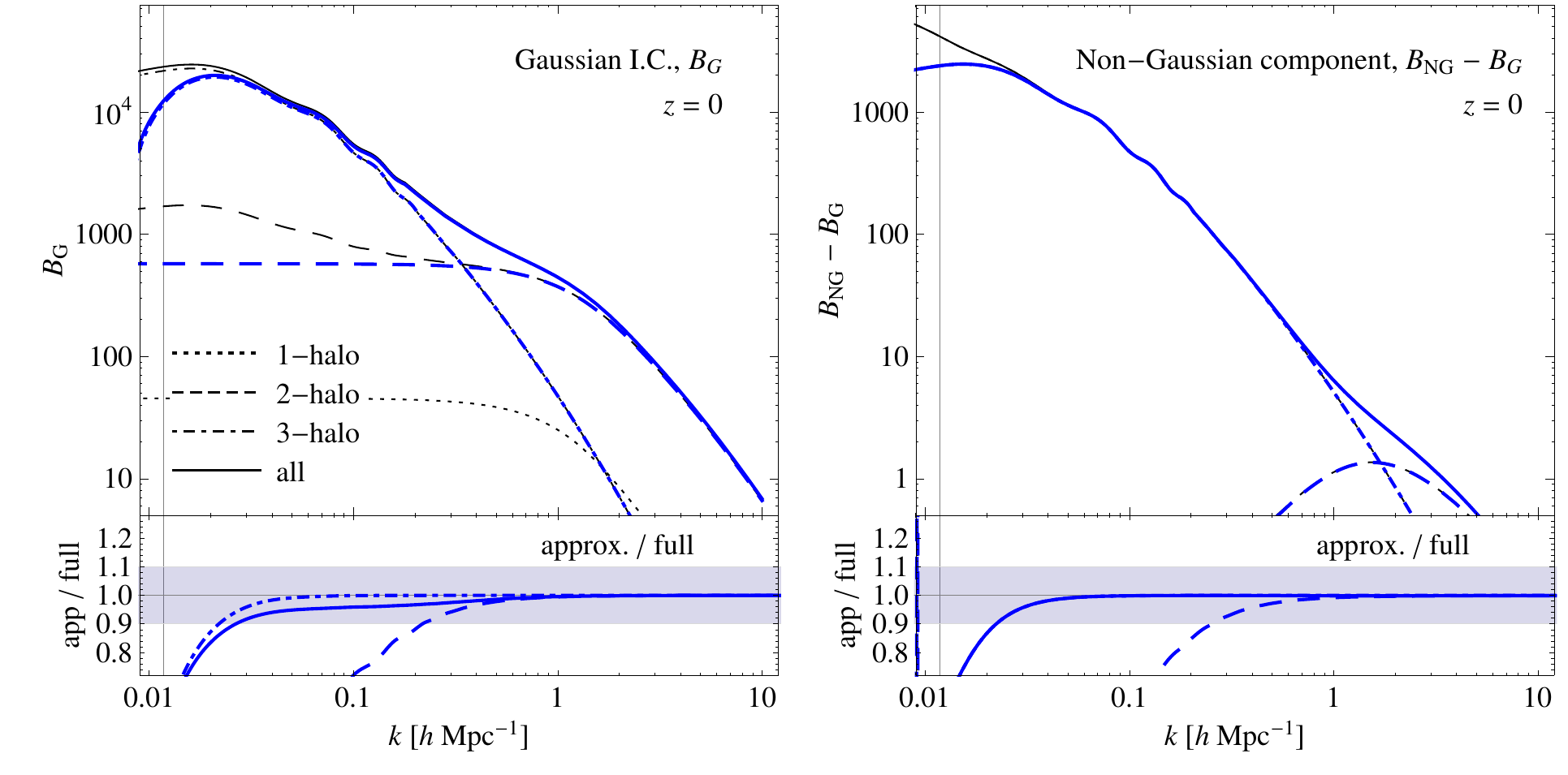}
\caption{Comparison between the full Halo Model prediction of Eqs.~(\ref{B3h})-(\ref{B1h}) ({\em thin, black curves}) and the approximate expressions of Eqs.~(\ref{pp1})-(\ref{pp3}) ({\em thick, blue curves}) for squeezed configurations with $k_1=0.014\kMpc$ as a function of $k_1=k_3=k$, at redshfit $z=0$. Dotted, dashed and dot-dashed curves correspond to the 1-, 2- and 3-halo contributions while the continuous curve correspond to their sum. On the left panel we show the Gaussian prediction $B_{\rm G}$ while on the right panel we show the non-Gaussian component $\Delta B_{\rm NG}\equiv B_{\rm NG}-B_{\rm G}$. The lower panels present the ratio between the approximations and the full expressions for each component. The shaded area marks value within a 10\% discrepancy. The thin vertical line indicates the value of $k_1$, and therefore corresponds effectively to an {\em equilateral} bispectrum configuration.}
\label{fig:app}
\end{figure}

As expected, the approximate predictions describe well the exact evaluations for large values of $k$. More precisely, the difference between the two is below 10\% for $k>0.02\kMpc$ with $k_1=0.014\kMpc$, that is, already for a ratio between the two sides given by $k/k_1\gtrsim 1.5$. For $k/k_1\gtrsim 10$ the error is at the percent level for $B_{\rm G}$ and even lower for $\Delta B_{\rm NG}$. However, limiting our attention to the approximation of the $B_{2h}$ term alone, we notice that one needs to consider at least $k/k_1\gtrsim 10$ to reach an accuracy below 10\%. As expected, the approximation for $B_{1h}$ works better than $0.01\%$ for all the momenta range of interest, both in the Gaussian as in the non-Gaussian case. 

To conclude, for $k_1 \lesssim 0.2 \kMpc$ and $k > k_1$, the total bispectrum in the squeezed-limit, $k_1 \ll k = k_2 \simeq k_3$, with Gaussian ($\fNL = 0$) or non-Gaussian ($\fNL \neq 0$) initial conditions, can be very well described by the expression
\begin{equation}
\begin{split}
\label{eq:Yet_simple_formula}
B(k_1,k,k)  = {1 \over \bar{\rho}}\,\epsilon^{[m]}_{2}(k,\fNL) %\hspace*{8.5cm}
%\nonumber\\
 %& 
& + \left\{\epsilon^{[b_1]}_2(k,\fNL) + 2\,P_{2h}(k,\fNL)\left[\frac{13}{14} + \left( \frac{4}{7} - \frac12 \frac{d \ln P_{L}}{d \ln k}\right) (\hat \bk_1 \cdot \hat \bk)^2 \right.  \right.\\
 & \left. \left. + \frac{\epsilon_{1}^{[b_{2}]}(k,\fNL)}{\epsilon_{1}^{[b_{1}]}(k,\fNL)} + \frac{2\,\fNL}{M(k_1,z)} \right] \right\}\,P_L(k_1)\,,
\end{split}
\end{equation}
to the $\sim 10\%$ and $\sim 1\%$ levels for $k/k_1 \gtrsim 1.5$ and $k/k_1 \gtrsim 10$, respectively. Since at small scales ($k > 0.2\kMpc$) and for squeezed configurations $B_{2h}$ dominates over $B_{1h}$ and $B_{3h}$, for $k > 6\kMpc$ we find that the total bispectrum is well described (to better than 1\% as compared to the full Halo Model) simply by 
\be
\label{eq:simple_formula}
B(k_1,k,k) \simeq {1 \over \bar{\rho}}\,\epsilon^{[m]}_{2}(k,\fNL) + \epsilon_2^{[b_1]}(k,\fNL)\,P_L(k_1) \,,
\ee
where we have included the term $\epsilon^{[m]}_{2}(k,\fNL)/\bar{\rho}$ (coming from the approximation of $B_{1h}$) to guarantee a accuracy better than $1\%$. Should we drop it and consider only the term $\epsilon_2^{[b_1]}(k,\fNL)\,P_L(k_1)$, then the Bispectrum would be captured at a $5\%$ level for $k \gtrsim 6 \kMpc$. Only for $k > 30 \kMpc$, $\epsilon_2^{[b_1]}(k,\fNL)\,P_L(k_1)$ alone describes, to better than $1\%$, the full matter bispectrum in the squeezed configuration. 

%%%%%%%%%%%%%%%%%%%%%%%%%%%%%%%%%%%%%%%%%%%%%%%%%%%%%%%%%%%%%%%%%%%%%%%%%%%%%%%%
\subsection{Equilateral and generic triangles}
\label{sec:eq}
%%%%%%%%%%%%%%%%%%%%%%%%%%%%%%%%%%%%%%%%%%%%%%%%%%%%%%%%%%%%%%%%%%%%%%%%%%%%%%%%

\begin{figure}[p]
\includegraphics[width=0.98\textwidth]{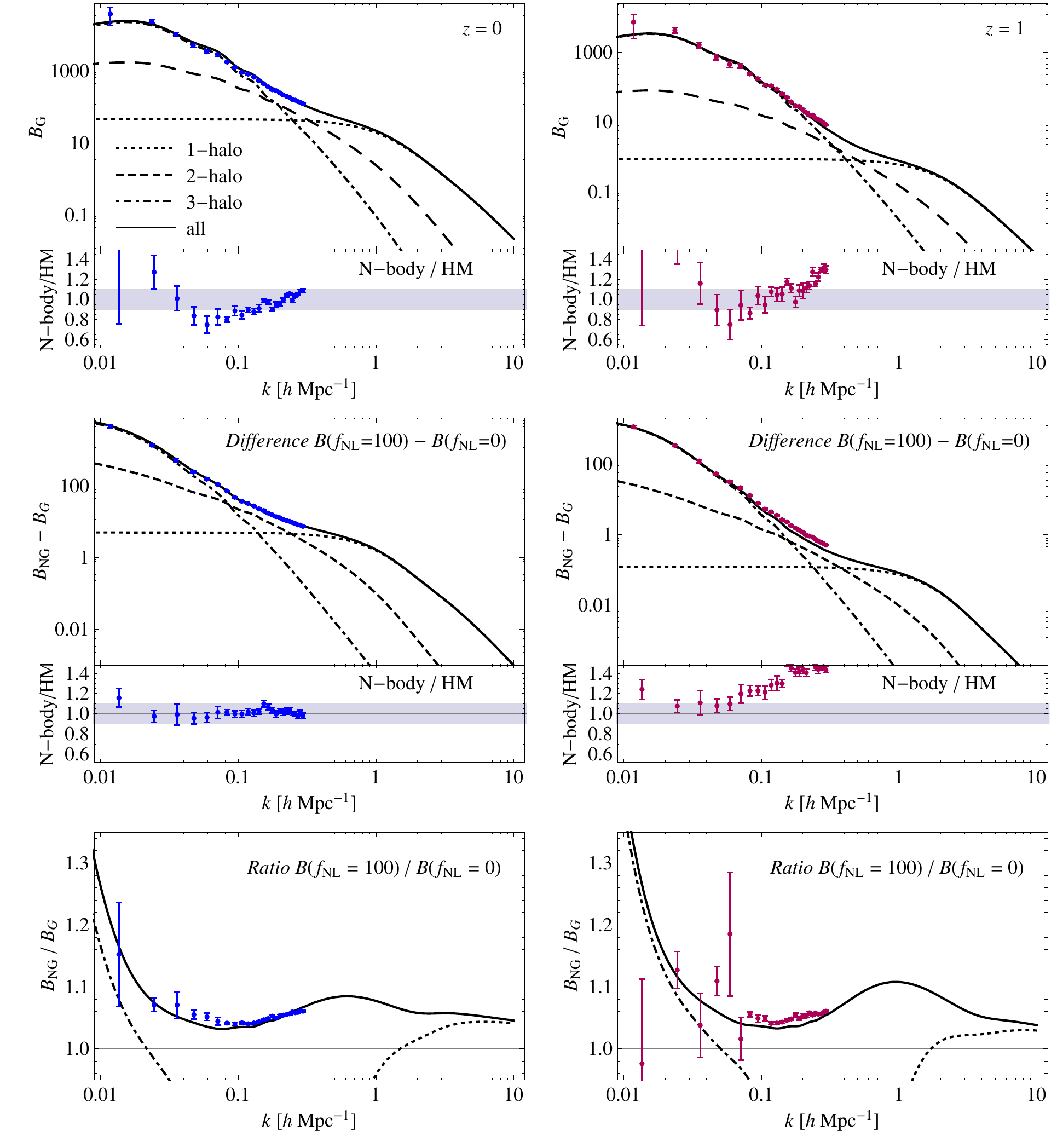}
\caption{
Equilateral configurations of the matter bispectrum, $B(k,k,k)$, at redshift $z = 0$ ({\em left panels}) and $z=1$ ({\em right panels}). See text for explanation.}
\label{fig:BsmEq}
\end{figure}
Figure~\ref{fig:BsmEq} shows the same results as in Fig.~\ref{fig:BsmSq} but for equilateral configurations, $k_1=k_2=k_3$. In particular, we plot  $B(k,k,k)$ as a function of $k \equiv k_1 =k_2 = k_3$ using the same line style as in Fig.~\ref{fig:BsmSq}. The first row shows the matter bispectrum for Gaussian initial conditions, together with the ratio between the measurements in N-body simulations and the model.  The second and third rows show, respectively, the difference and the ratio between the non-Gaussian and Gaussian case. For $z=0$ the agreement is reasonably good, especially on small scales. However, for $z=1$ the agreement worsens and for both redshift it does not remain within the 10\% accuracy found in the squeezed limit. On the other hand, it is reasonable to expect that such large discrepancy is limited only to the mildly nonlinear regime, where the HM typically fails. 

The non-Gaussian components represent a correction of a few up to $\sim 10$\%. An enhancement, shown as a bump in the ratio between the non-Gaussian and Gaussian bispectrum, is also displayed for equilateral configurations. Such local maximum is located at similar but slightly bigger scales w.r.t.~the squeezed case. We find it at $k \approx 0.65 \kMpc$ for $z = 0$ and at $k \approx 1.0 \kMpc$ for $z = 1$. The difference in the correction when comparing the bispectrum for $z = 0$ versus the one for $z = 1$, is not as big as in the squeezed configuration, where there was a change from a $15\%$ to $25\%$ enhacement. Here we rather observe a change in the correction from $9\%$ at $z = 0$ to $11\%$ at $z = 1$.

The ratio of the difference $B_{\rm NG}-B_{\rm G}$ between the N-body simulations and our predictions is inside the 10\% agreement for $z=0$, but largely outside for $z=1$. We do not attempt here to explain why the Halo Model is less accurate for equilateral configurations when local non-Gaussianity is considered. However, since  the ratio $B_{\rm NG}/B_{\rm G}$ agrees very well with the one measured in the numerically simulations, this inaccuracy is likely to be traced in the Halo Model itself, more than on the inclusion of non-Gaussian corrections.

In order to provide a more complete description of the impact of primordial non-Gaussianity on all triangular configurations, in Fig.~\ref{fig:BsmAll} we show  the relative effect $B_{\rm NG}(k_1,k_2,k_3)/B_{\rm G}(k_1,k_2,k_3)$ at $z=1$ as a function of the ratios $k_3/k_1$ and $k_2/k_1$ assuming a constant $k_1=3 \kMpc$. In order to avoid redundancy among equivalent configurations, the quantity plotted takes values on a triangle (represented by the shaded area on the bottom surface) where the lower corner corresponds to flattened triangles ($k_1=2k_2=2k_3$), the top right corner to equilateral configurations ($k_1=k_2=k_3$) and the top left corner to squeezed configurations ($k_2\ll k_1=k_3$). The left panel shows the ratio $B_{\rm NG}/B_{\rm G}$ computed in the tree-level approximation in PT. Even though perturbation theory is not applicable at these scales, this plot nevertheless provides an estimate of the contribution of the linear bispectrum $B_0$ to the overall effect. Clearly, a large effect is 
present only in the squeezed limit. The right panel shows instead the full HM calculation. In this case we notice how, in addition to a 20\% effect in the squeezed limit, {\em all} triangles of essentially any shape receive a correction of the order of about 7-8\%. We also notice, as in the previous plots, a local maximum effect corresponding to wavenumbers of order $1\kMpc$ and to flattened configurations.

\begin{figure}[t]
\includegraphics[width=0.48\textwidth]{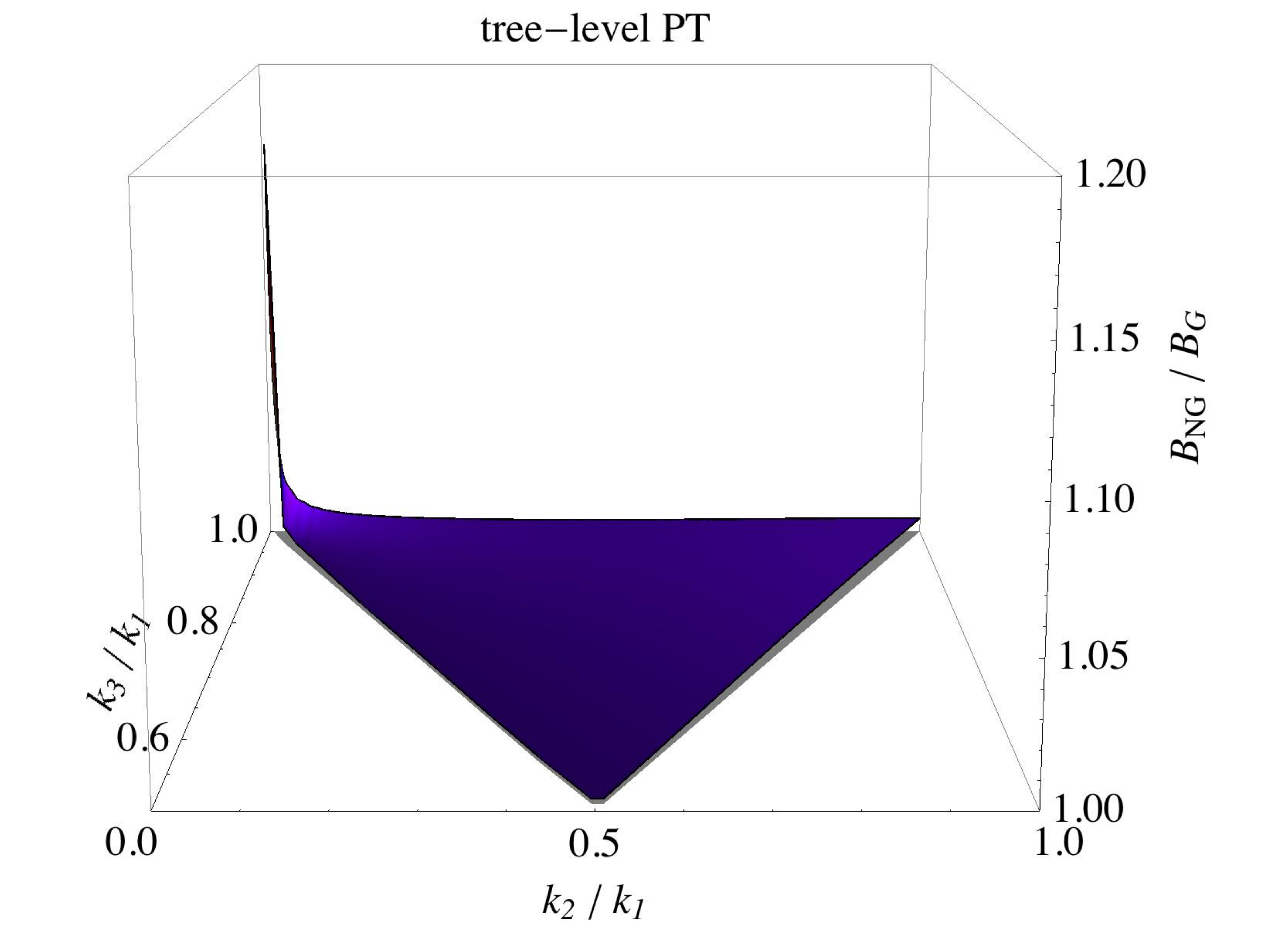}
\includegraphics[width=0.48\textwidth]{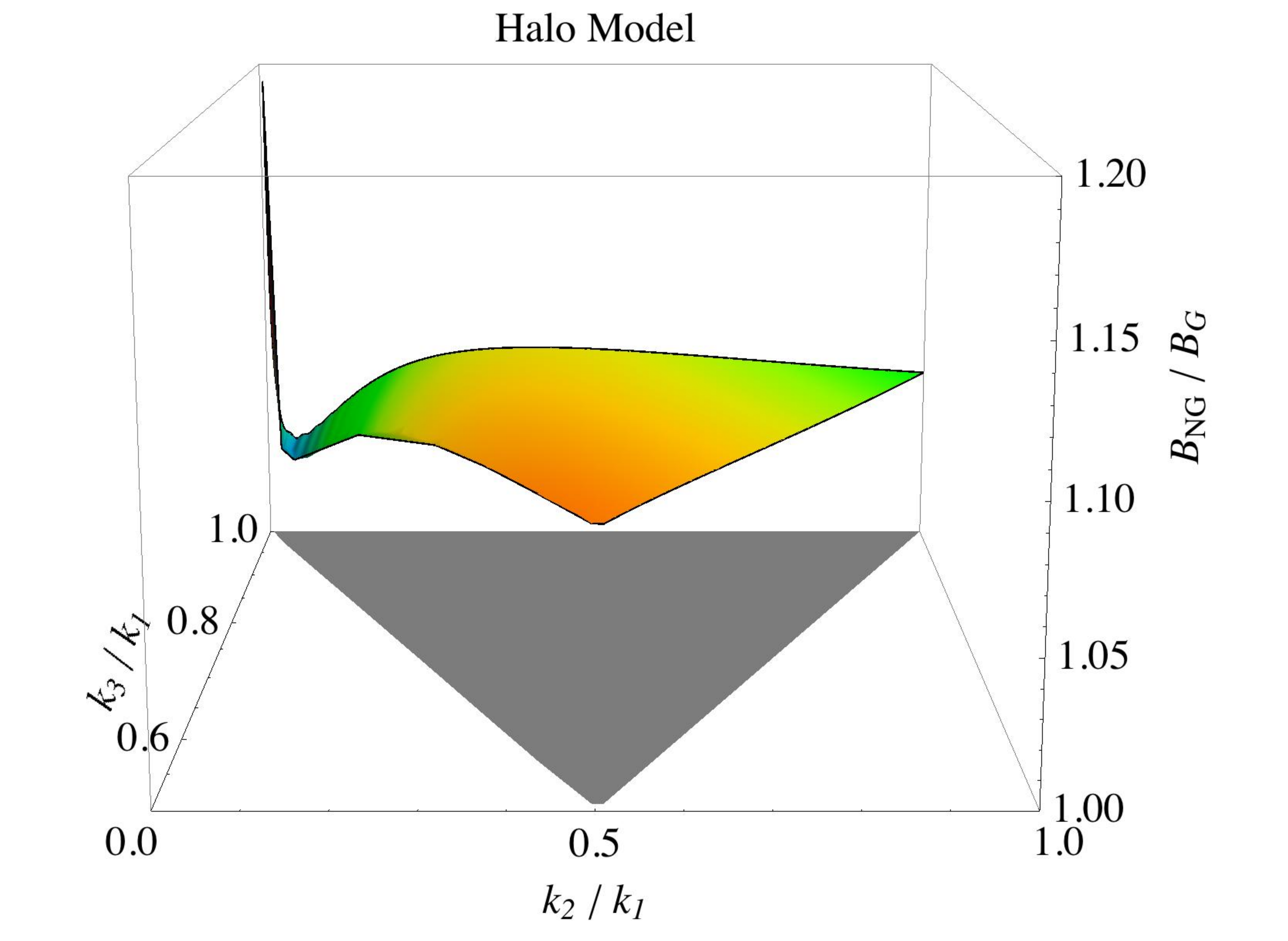}
\caption{Relative effect of primordial non-Gaussianity shown in terms of the quantity $B_{\rm NG}(k_1,k_2,k_3)/B_{\rm G}(k_1,k_2,k_3)$ at $z=1$ plotted as a function of the ratios $k_3/k_1$ and $k_2/k_1$ assuming a constant $k_1=3 \kMpc$. The function takes values on a triangle (represented by the shaded area on the bottom surface) where the lower corner corresponds to flattened triangles ($k_1=2k_2=2k_3$), to top right corner to equilateral configurations ($k_1=k_2=k_3$) and the top left corner to squeezed configurations ($k_2\ll k_1=k_3$). The left panel shows the ratio $B_{\rm NG}/B_{\rm G}$ computed in the three-level approximation in PT. The right panel shows instead the full HM calculation. }
\label{fig:BsmAll}
\end{figure}

%%%%%%%%%%%%%%%%%%%%%%%%%%%%%%%%%%%%%%%%%%%%%%%%%%%%%%%%%%%%%%%%%%%%%%%%%%%%
\section{Conclusions}
\label{sec:conclusions}
%%%%%%%%%%%%%%%%%%%%%%%%%%%%%%%%%%%%%%%%%%%%%%%%%%%%%%%%%%%%%%%%%%%%%%%%%%%%

In this paper we have used the Halo Model approach to compute, over a wide range of scales including very small ones, the matter bispectrum for non-Gaussianity of the local type. We have considered non-Gaussian corrections which appear in the three fundamental ingredients of the Halo Model: the halo profile, the  halo mass function and the bias functions. For the Gaussian case we have used the NFW \citep{NavarroFrenkWhite1997} halo profile and the Sheth \& Tormen~\citep[ST,][]{ShethTormen1999} expressions for the halo mass and bias functions. The non-Gaussian correction to the halo profile has been included by using \eq{eq:rhoNGb}, 
which is derived as a fit to the numerical simulations presented in \citet{SmithDesjacquesMarian2011}. For the correction to the halo mass function we have employed
the expression proposed by \citet{LoVerdeEtal2008} based on the Edgeworth expansion about the Press-Schechter mass function, with $\nu = {\delta_c}/{\sigma}$ rescaled by a  scaling parameter $q=0.91$. Finally, for the bias functions we have assumed a simplified version of the expressions in \citet{BaldaufSeljakSenatore2011} and \citet{SefusattiCrocceDesjacques2011}, as described in sec.~\ref{ssec:bias}.

Using these ingredients, we have computed the matter power spectrum and bispectrum, both in the Gaussian and non-Gaussian cases. For a local parameter $f_{\rm NL}= 100$, in \citet{FedeliMoscardini2010} and \citet{SmithDesjacquesMarian2011} it was noted that non-Gaussian corrections enhance the matter power spectrum on scales $k \sim 1 h^{-1}$Mpc, inducing a bump in the ratio between the non-Gaussian and Gaussian power spectrum. As shown in Fig.~\ref{fig:breakdown}, this bump is mainly due to the corrections to the halo mass function, and represents a $2-3\%$ correction. For the same parameter $f_{\rm NL}= 100$, we have shown that the non-Gaussianity introduces a correction in the squeezed configuration $k_1 \ll k_2 \simeq k_3 = k$ of the bispectrum, of about $\sim$ 15\% for redshift $z = 0$ and $\sim$ 25\% for redshift $z = 1$, also at $k \sim 1 \kMpc$. We thus confirm the presence of a similar bump also in the ratio between the non-Gaussian and Gaussian bispectra. We have also shown that the non-Gaussian 
corrections in the equilateral configuration $k_1 = k_2 = k_3 = k$ of the bispectrum, reach a maximum at $k \sim 1 \kMpc$ of about $\sim$ 10\%, both for redshift $z = 0$ and $z = 1$. The corrections due to local non-Gaussianity in the equilateral configuration if the bispectrum are thus less sensitive to the redshift.

We have also studied how the non-Gaussian corrections affect each of the halo model ingredients individually, which is shown in Fig.~\ref{fig:breakdown}. This has allowed us to conclude that, contrary to the case of the matter power spectrum, it is the non-Gaussian correction in the linear bias which determines the shape and amplitude of the maximum distortion of the squeezed configuration of the Bispectrum, in the presence of local non-Gaussianity. 
Moreover, the corrections to the halo profiles are subdominant in the bispectrum over all the scales considered in this paper. Finally, we conclude that all corrections of the Halo Model ingredients might be important, since different correlators seems to be sensitive to different corrections of the ingredients.

We have also compared our results to numerical simulations both for squeezed and equilateral configurations. In the squeezed limit the agreement is remarkably good, both in the Gaussian and non-Gaussian case. In particular, it is better than about 10\% for $z=0$, both for the total bispectrum and for its non-Gaussian correction. This strongly suggests that in the squeezed limit the Halo Model predictions for  the matter bispectrum work reasonably well and they capture with very good accuracy non-Gaussian corrections of the local type. For equilateral configurations the agreement between the Halo Model and N-body simulations is still reasonably good for $z=0$ but worsens for $z=1$.   

In real space, the squeezed limit represents the case where two points are close to each other, belonging to the same halo, and far from the third point, belonging to a different halo. In this limit we have checked that the bispectrum is correctly dominated by the 2-halo term. Moreover, on large scales the Fourier transform of the halo profile in the long mode can be simply approximated as the mass of the halo. This allows a considerable simplification of the Halo Model expressions, which can be described by very simple formulas, \eq{eq:Yet_simple_formula} at mildly non-linear scales and \eq{eq:simple_formula} at highly non-linear scales. These represent master formulas for the matter bispectrum in the squeezed configuration independently of the nature of the initial conditions. 

Given the agreement that we have found between the Halo Model calculation and the simulations for the matter bispectrum, it seems promising to use the Halo Model to compute observable quantities such as the galaxy or lensing bispectra. We indeed provide a preliminar estimate of the galaxy bispectrum in the presence of local non-Gaussianity in Appendix~\ref{sec:gbisp}.

\subsection*{Acknowledgements}
D.~G.~F.~is supported by the Academy of Finland grant 131454 and also acknowledges support from a Marie Curie Early Stage Research Training Fellowship associated with the EU RTN 'UniverseNet', during his stay at CERN TH-Division when this project was initiated. A.R. is supported by the Swiss National Science Foundation (SNSF), project `The non-Gaussian Universe" (project number: 200021$_{}$140236). E.S. was supported in part by the EU Marie Curie Inter-European Fellowship.

\appendix
%%%%%%%%%%%%%%%%%%%%%%%%%%%%%%%%%%%%%%%%%%%%%%%%%%%%%%%%%%%%%%%%%%%%%%%%
\section{The Galaxy Bispectrum}
\label{sec:gbisp}
%%%%%%%%%%%%%%%%%%%%%%%%%%%%%%%%%%%%%%%%%%%%%%%%%%%%%%%%%%%%%%%%%%%%%%%%%%%%

The Halo Model framework can be extended to predict galaxy correlators by means of the Halo Occupation Distribution (HOD). In fact, the additional ingredient required is given by a  prescription describing the way galaxies populate dark matter halos. In analogy with \eq{eq:HaloModelPowerSpectrum}, we could write the galaxy power spectrum as the sum
\citep{Seljak2000, ScoccimarroEtal2001A}

\begin{equation}
P_{\rm g}(k)=P_{1h,g}(k)+P_{2h,g}(k),
\end{equation}
where now the two contributions are given by
\begin{eqnarray}
\label{pdg}
P_{1h,{\rm g}}(k) & = & \int\!dm\, n(m)\,\frac{\langle N_{\rm g}(N_{\rm g}-1)\rangle_m}{\bar{n}_{\rm g}^2}\,\frac{{\hat\rho}^2(k,m)}{m^2}\,,\\
P_{2h,{\rm g}}(k) & = & \left[\prod_{i=1}^2\int\! dm_i\,n(m_i)\,\frac{\langle N_{\rm g} \rangle_{m_i}}{\bar{n}_{\rm g}}\,\frac{{\hat\rho}(k,m_i)}{m_i}\right] P_h(k,m_1,m_2)\,,
\end{eqnarray}
with $\bar{n}_{\rm g}$ being the mean galaxy number density and where $\langle N_{\rm g} \rangle_{m}$ and $\langle N_{\rm g}(N_{\rm g}-1)\rangle_m$ represent, respectively, the first and second moments of the probability $P(N|m)$ for a halo of mass $m$ to contain $N$ galaxies. In simple terms, $\langle N_{\rm g} \rangle_{m}$ describes the mean number of galaxies in halos of mass $m$ while $\langle N_{\rm g}(N_{\rm g}-1)\rangle_m$ describes instead the mean number of galaxy couples in the same halo. Similarly, for the bispectrum we have
\begin{equation}
B_{\rm g}(k_1,k_2,k_3)=B_{1h,g}(k_1,k_2,k_3)+B_{2h,g}(k_1,k_2,k_3)+B_{3h,g}(k_1,k_2,k_3),
\end{equation}
with
\begin{eqnarray}\label{eq:bg1h}
B_{1h,{\rm g}}(k_1,k_2,k_3) & = & \int\! dm\, n(m)\,\frac{\langle N_{\rm g}(N_{\rm g}-1)(N_{\rm g}-2)\rangle_m}{\bar{n}_{\rm g}^3}\,\frac{\hat\rho(k_1,m)\,\hat\rho(k_2,m)\,\hat\rho(k_3,m)}{m^3}\,,\\
\label{eq:bg2h}
B_{2h,{\rm g}}(k_1,k_2,k_3) & = & \int\! dm'\, n(m')\,\frac{\langle N_{\rm g}(N_{\rm g}-1) \rangle_{m'}}{\bar{n}_{\rm g}^2}\,\frac{\hat\rho(m',k_2)\,\hat\rho(m',k_3)}{m'^2}\\
 & & \times \int\! dm\,n(m)\,\frac{\langle N_{\rm g} \rangle_{m}}{\bar{n}_{\rm g}}\,\frac{\hat\rho(m,k_1)}{m}\,P_h(k_1,m,m')+{\rm 2~perm.}\,,\nonumber\\
\label{eq:bg3h}
B_{3h,{\rm g}}(k_1,k_2,k_3) & = & \left[\prod_{i=1}^3\int\!dm_i\,n(m_i)\,\frac{\langle N_{\rm g} \rangle_{m_i}}{\bar{n}_{\rm g}}\,\frac{\hat\rho(m_i,k_i)}{m_i}\right] B_h(k_1,m_1;k_2,m_2;k_3,m_3)\,,
\end{eqnarray}
where, in addition, the 1-halo term now requires the third moment\footnote{Following the review of \citet{CooraySheth2002} one should notice that: 1) in principle the spatial distribution of galaxies inside the halo does not necessarily follow the dark matter distribution, but it is indeed a good approximation; 2) if an halo contains a single galaxy then this is expected to sit at the center of the halo itself, and the double power of the profile should be replace by a single one. These details are not essential for our purposes.} of the HOD, $\langle N_{\rm g}(N_{\rm g}-1)(N_{\rm g}-2)\rangle_m$.

A given HOD clearly depends on the specific galaxy population that one wants to describe. In the following, for illustration purposes we will consider the case of Luminous Red Galaxies (LRG), since this is today the best know galaxy type. It has been shown that a proper description of the LRGs HOD can be obtained by separately considering, for halos containing more than one galaxy, a {\em central} galaxy ({\em i.e.} a single galaxy placed at the halo center) and a population of {\em satellite} galaxies following a Poisson distribution \citep{KravtsovEtal2004}. In particular we will adopt the following parametrization for the mean
\begin{equation}
\langle N_{\rm g}\rangle_m=\langle N_c\rangle_m+\langle N_s\rangle_m\,,
\end{equation}
with
\bea
\langle N_c\rangle_m & = & \exp\left(-\frac{m_{\rm min}}{m}\right)\,,\\
\langle N_s\rangle_m & = & \exp\left(-\frac{m_{\rm min}}{m}\right)\left(\frac{m}{m_1}\right)^\alpha\,,
\eea
where $m_{\rm min}$ represents the mass threshold above which we expect at least one, central galaxy while $m_1$ is the typical mass above which a second galaxy is present. We assume the values of the three parameters $m_{\rm min}$, $m_1$ and $\alpha$ from \citet{KulkarniEtal2007}, which obtain them from a fit to measurements of the LRGs three-point function in the Sloan Digital Sky Survey.
Their best fit corresponds to $m_{\rm min}=7.66\times 10^{13}\Ms$, $m_1=4.7\times 10^{14}\Ms$ and $\alpha=1.4$. 
In order to compute the power spectrum and bispectrum 1-halo terms we will need as well the second and third order moments. Since we assume the satellite HOD to correspond to a Poisson distribution, then we simply have 
\be
\langle N_s(N_s-1)...(N_s-j)\rangle_m=\langle N_s\rangle_m^{j+1}
\ee
so that
\bea
\langle N_{\rm g}(N_{\rm g}-1)\rangle_m & = & \langle N_s\rangle_m\left(\langle N_s\rangle_m+2\right),\\
\langle N_{\rm g}(N_{\rm g}-1)(N_{\rm g}-2)\rangle_m & = &  \langle N_s\rangle_m^2\left(\langle N_s\rangle_m+3\right).
\eea

In these calculations, the problem represented by the constant value assumed by the 2-halo and 1-halo contributions at small $k$ is more severe than in the matter case. In fact, for both squeezed or equilateral configurations such contributions are not negligible at large scales. A possible solution to this problem has been proposed by assuming ``compenpensated profiles'' for the halo matter ditribution \citep[see, {\em e.g.}, section 4.4 in][]{CooraySheth2002}. In our illustrative calculations we adopt the simple prescription consisting in replacing the NFW profile $\hat{\rho}(m,k)$ with the difference $\hat{\rho}(m,k)-m W_R(k)$, $W_R(k)$ being the top-hat window function of radius $R  = [3 m/(4\pi \bar\rho)]^{1/3}$. We perform such replacement only for the profiles with $k_2 \approx k_3 = k$ in the expression for the 1-halo contribution, Eq.~(\ref{eq:bg1h}), and for those in the first integral in the 2-halo contribution in Eq.~(\ref{eq:bg2h}). While this is clearly not a well justified procedure, our 
intent is to avoid a largely incorrect behaviour at large scales and focus on the small scale non-Gaussian correction, which is not affected by this choices.

\begin{figure}[p]
\includegraphics[width=0.98\textwidth]{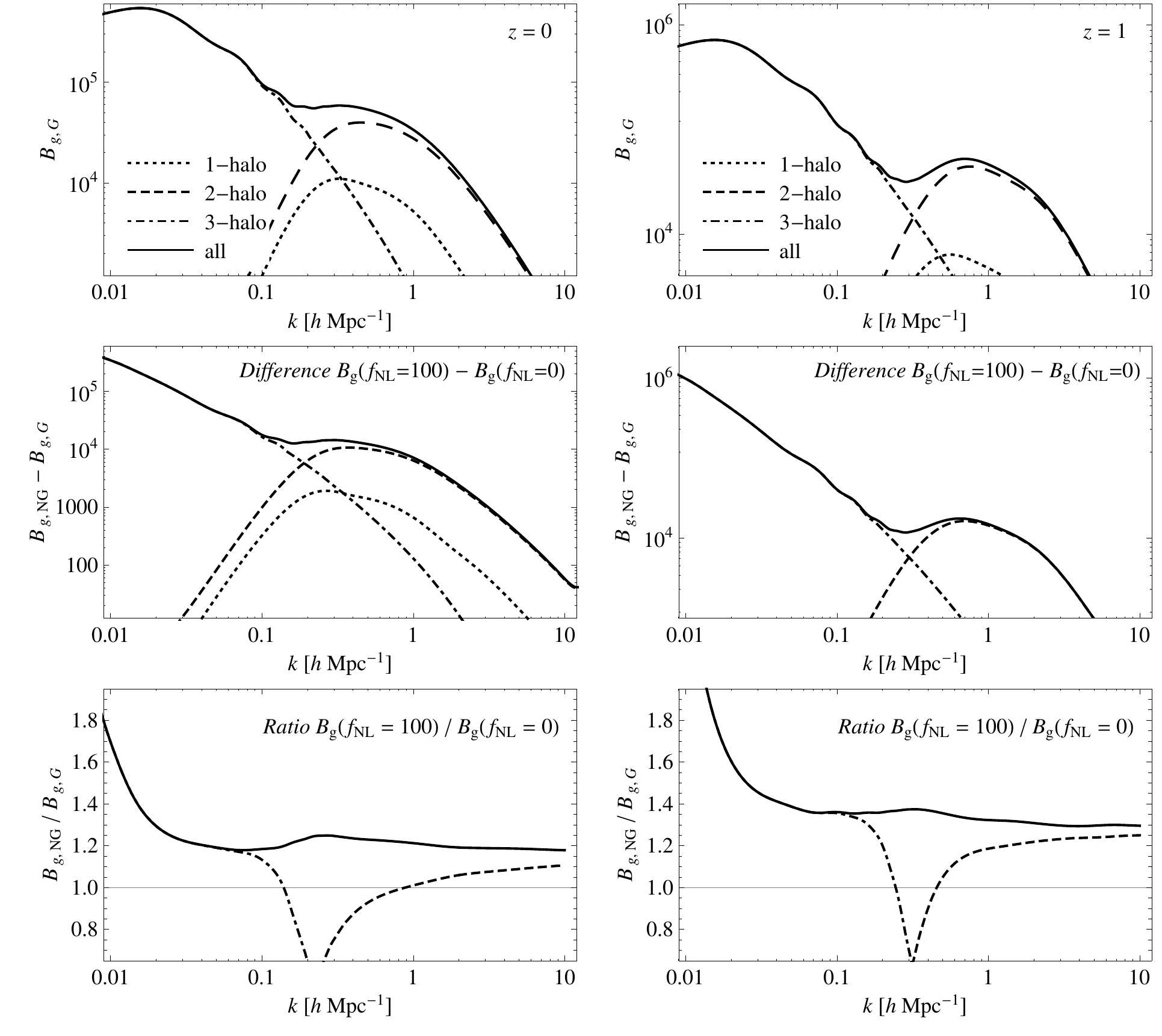}
\caption{
Squeezed configurations of the galaxy bispectrum, $B_{\rm g}(k_1,k,k)$, with $k_1 \ll k_2 = k_3 \equiv k$, at redshift $z = 0$ ({\em left panels}) and $z=1$ ({\em right panels}). See text for explanation.
}
\label{fig:BsgSq}
\end{figure}
In Fig.~\ref{fig:BsgSq} we show the results for squeezed configurations of the galaxy bispectrum $B_{{\rm g}}(k_1,k,k)$, defined by $k_1=0.014\kMpc$ fixed, plotted as a function of $k_2 = k_3 \equiv k$. Similarly to the matter case, left panels show the results at redshift $z=0$, while right panels at $z=1$. In the first row we present the galaxy bispectrum for Gaussian initial conditions. The middle panels show the component due to non-Gaussian initial conditions with a local $\fNL=100$, that is the {\em difference} between the predictions with and without non-Gaussianity, $\Delta B_{\rm g}^{\rm NG}\equiv B_{\rm g}^{\rm NG}(\fNL=100)-B_{\rm g}^{\rm G}(\fNL=0)$ while the lower panels show the non-Gaussian to Gaussian ratio. In all plots, the dotted, dashed and dot-dashed curves correspond, respectively, to the 1-halo, 2-halo and 3-halo contributions, while the continuous line corresponds to their sum. 

Again the dominant term in this squeezed limit  is given by $B_{2h}(k_1,k,k)$. It is interesting to notice how, despite presenting as well a maximum around $k\simeq 1\kMpc$ at $z=0$, unlike in the matter case, the ratio between the non-Gaussian galaxy bispectrum to the Gaussian one, shows a more constant behaviour as $k$ grows. The non-Gaussian effect is, in fact, of the order of 20\% (30\%) over a large range of scales at redshift $z = 0$ ($z = 1$). %20\% over a large range of scales both at low and high redshift.
This is a significant correction that might be taken into account in studies of galaxies bias at small scales. 

As in the matter case, for the galaxy bispectrum as well we can find simplified expressions for squeezed triangular configurations. The constraints of \eq{eq:condrho} and (\ref{eq:condbias}), will be now modified into
\bea
\label{ndmGal}
{1\over\bar{n}_{\rm g}}\int\! dm\, n(m)\,\langle N_{\rm g} \rangle_m\,b_1(m,k) &\equiv& b_{1,{\rm g}}(k)\,, \\
\label{ndm2Gal}
{1\over\bar{n}_{\rm g}}\int\! dm\, n(m)\,\langle N_{\rm g} \rangle_m\,b_2(m,k) &\equiv& b_{2,{\rm g}}(k)\,,
\eea
where $b_{1,{\rm g}}$ and $b_{2,{\rm g}}$ represent the linear and quadratic bias of the galaxy population. Taking into account these relations we can obtain now a squeezed approximation for the galaxy bispectrum, using, as we did in the matter case, the fact that $\hat\rho(k) \rightarrow m$ when $k \rightarrow 0$. For $k_1 \ll k$ we find
\begin{align}
\label{eq:bg1happ}
B_{1h,g}(k_1,k,k) &= {1 \over \bar{n}_{\rm g}}\,\epsilon^{[\langle N_{\rm g}(N_{\rm g}-1)(N_{\rm g}-2)\rangle]}_{g,2}(k,f_{\rm NL})\,,\\
\label{eq:bg2happ}
B_{2h,g}(k_1,k,k) &=
b_{1,{\rm g}}(k_1)\,\epsilon_{g,2}^{[b_1\langle N_{\rm g}(N_{\rm g}-1)\rangle]}(k,f_{\rm NL})\,P_L(k_1)\,,\\
\label{eq:bg3happ}
B_{3h,g}(k_1,k,k) &= 2\,b_{1,{\rm g}}(k_1)\,\left[ \frac{13}{14} + \left( \frac{4}{7} - \frac12 \frac{d \ln P_{L}}{d \ln k}\right) (\hat \bk_1 \cdot \hat \bk_2)^2 + \frac{\epsilon_{g,1}^{[b_2\langle N_{\rm g}\rangle]}(k,\fNL)}{\epsilon_{g,1}^{[b_1\langle N_{\rm g}\rangle]}(k,\fNL)} + \frac{2\,\fNL}{M(k_1)} \right]\,P_L(k_1)P_{2h,g}(k)\,,
\end{align}
where the $\epsilon_{g,i}^{[F]}$ functions read now
\be
 \epsilon_{g,i}^{[F]}(k,\fNL) \equiv \frac{1}{\bar{n}_{\rm g}^i}\int\! dm\,n(m,\fNL)\left[{\hat{\rho}(k,m,\fNL)\over m}\right]^i\,F(m,\fNL)\,.
\ee
\begin{figure}[t]
\includegraphics[width=0.98\textwidth]{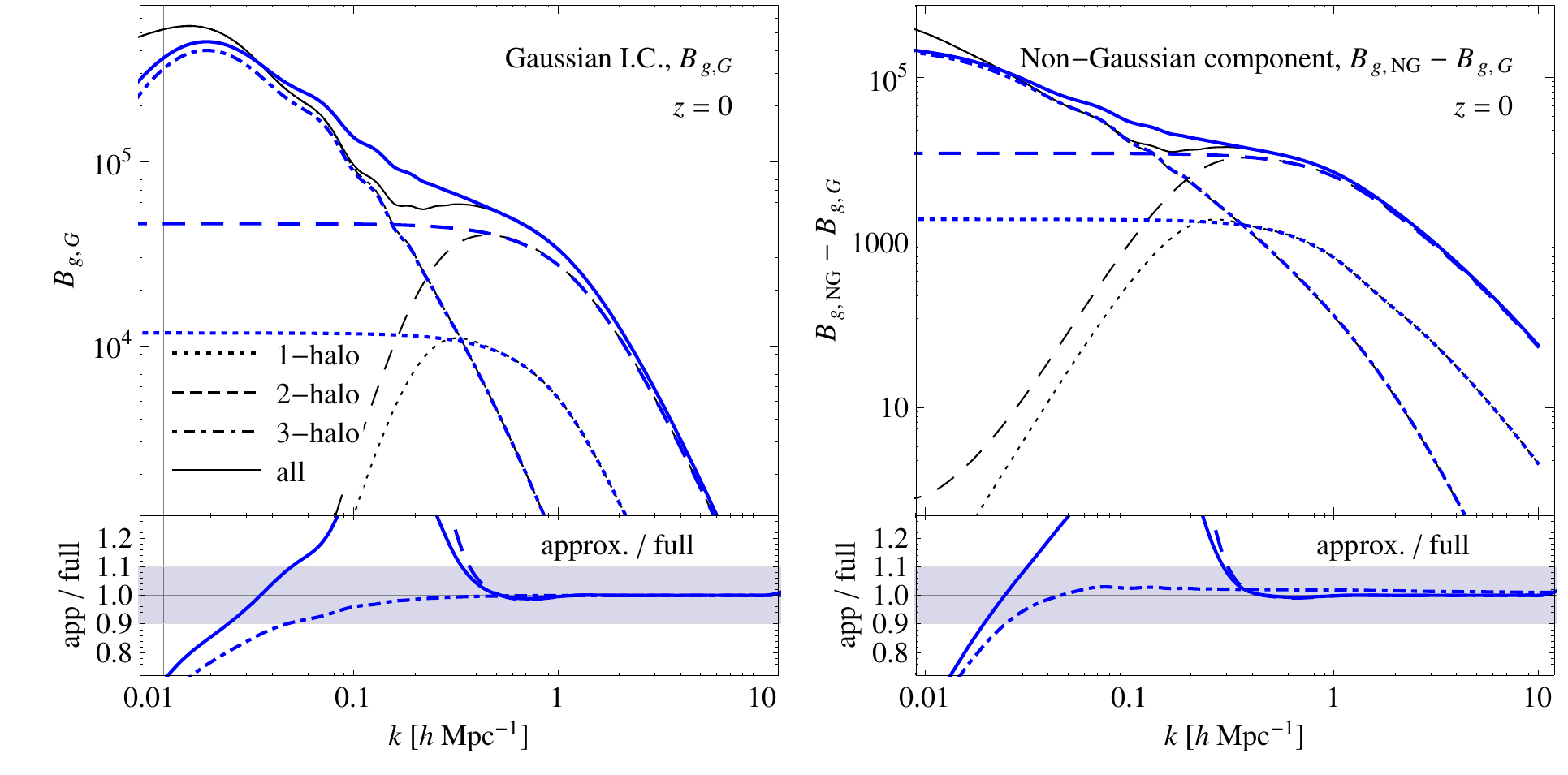}
\caption{Comparison between the full Halo Model prediction of the galaxy bispectrum of Eqs.~(\ref{eq:bg1h}), (\ref{eq:bg2h}) and (\ref{eq:bg3h}) ({\em thin, black curves}) and the approximate expressions of Eqs.~(\ref{eq:bg1happ}), (\ref{eq:bg2happ}) and (\ref{eq:bg1happ}) ({\em thick, blue curves}) for squeezed configurations with $k_1=0.014\kMpc$ as a function of $k_1=k_3=k$, at redshfit $z=0$. Dotted, dashed and dot-dashed curves correspond to the 1-, 2- and 3-halo contributions while the continuous curve corresponds to their sum. On the left panel we show the Gaussian prediction $B_{g,G}$ while on the right panel we show the non-Gaussian component $\Delta B_{g,NG}\equiv B_{g,NG}-B_{g,G}$. The lower panels present the ratio between the approximations and the full expressions for each component. The shaded area marks value within a 10\% discrepancy. The thin vertical line indicates the value of $k_1$, and therefore corresponds effectively to an {\em equilateral} bispectrum configuration.}
\label{fig:appgal}
\end{figure}

%Like in the matter case, for squeezed isosceles configurations $k_1 \ll k_2 = k_3 \equiv k$, one can simply drop the term $\propto (\hat \bk_1 \cdot \hat \bk_2)^2$ from the bracket of Eq.~(\ref{eq:bg3happ}), since it is just of order $\mathcal{O}(k_1/k)^2 \ll 1$. 
In Fig.~\ref{fig:appgal} we compare the simplified expressions Eqs.~(\ref{eq:bg1happ})--(\ref{eq:bg3happ}) versus the exact Halo Model evaluation, Eqs.~(\ref{eq:bg1h})--(\ref{eq:bg3h}), in the case of local NG with $f_{\rm NL}^{\rm loc} = +100$. The left panels consider the case of Gaussian initial conditions while the right panels show the non-Gaussian difference, as done in Fig.~\ref{fig:app} for the matter bispectrum. It is evident that in order to achieve an agreement better than 10\% it is necessary, in all cases and for every individual term, to consider $k \gtrsim 0.5\kMpc$ for $k_1=0.014\kMpc$, that is a ratio between the two sides of $k/k_1\gtrsim 40$. For $k \gtrsim 1 \kMpc$ ($k/k_1\gtrsim 100$), the simplified expressions reproduce the full ones at better than $1\%$ level. 

%%%%%%%%%%%%%%%%%%%%%%%%%%%%%%%%%%%%%%%%%%%%%%%%%%%%%%%%%%%%%%%%%%%%%%%%
\section{Details on the mass integrals}
\label{sec:AppendixB}
%%%%%%%%%%%%%%%%%%%%%%%%%%%%%%%%%%%%%%%%%%%%%%%%%%%%%%%%%%%%%%%%%%%%%%%%%%%%

\noindent
In this work all mass integrals in the HM expressions are performed numerically. An appropriate upper bound to such integration is easy to find, since the halo mass function decays exponentially fast. On the other hand it is clearly not feasible to set to zero the integral's lower bound. We are then forced to cut the integration at some minimum value $M_{\rm min}$ and compensate for the distribution low-mass tail removed for $M < M_{\rm min}$. In this Appendix we explain our prescription for this procedure and show that, for sufficiently small values of $M_{\rm min}$, our results are insensitive to such cut-off scale, as it should be. This is particularly important for the constraint equations, Eqs.~(\ref{eq:condrho})-(\ref{eq:condbias}). To keep general the discussion we will then discuss the non-Gaussian case, while the Gaussian case can simply be obtained by setting $f_{\rm NL} = 0$.

Let us begin by the conservation of mass contraint equation,
\be\label{eq:condrhoNG_v2}
\frac{1}{\bar{\rho}} \int \!\! dM\,M\,n(M,z;\fNL) = 1\,.
\ee
Since we will only integrate over a finite range $[M_{\rm min},M_{\rm max}]$, we can compensate for the removed low mass tail distribution at $M < M_{\rm min}$, by introducing a constant $C_0$, such that
\be\label{eq:condrhoNG_v3}
\frac{1}{\bar{\rho}}\int \!\! dM\,M\,n(M,z;\fNL) = \frac{1}{\bar{\rho}}\int_{M_{\rm min}}^{M_{\rm max}} \!\! dM\,M\,n(M,z;\fNL) + C_0\frac{M_{\rm min}}{\bar{\rho}} = 1.
\ee
This is equivalent to consider the mass function, for $M \leqslant M_{\rm min}$, as a Dirac-delta like $n(M,z;\fNL) = C_0(z,\fNL;M_{\rm min})\delta_D(M-M_{\rm min})$. The constant $C_0$ can be easily obtained from the above equation, as
\bea
C_0(z,\fNL;M_{\rm min}) = [1-D_0(z,\fNL;M_{\rm min})]\frac{\bar{\rho}}{M_{\rm min}}\,,\\
D_0(z,\fNL;M_{\rm min}) \equiv \frac{1}{\bar{\rho}}\int_{M_{\rm min}}^{M_{\rm max}} \!\! dM\,M\,n(M,z;\fNL).
\eea 
For any choice of $M_{\rm min}$ there will be always a constant $C_0$, given in terms of $D_0$, that will just act in such a manner that the mass contraint will be attained like in \eq{eq:condrhoNG_v3}. The constant $D_0$ measures the deviation from unity when the constraint equation is evaluated by integrating within $[M_{\rm min},M_{\rm max}]$. Theorefore, for any integration over the mass in the Halo Model expressions involving the product of the mass function $n(M)$ times some other function $F(M)$ (except the bias functions, see below), we will simply apply the prescription
\bea\label{eq:IRmassCorrForMassFunct}
\int \!\! dM\,n(M,z;\fNL)\,F(M;...) & = & \int_{M_{\rm min}}^{M_{\rm max}} \!\! dM\,M\,n(M,z;\fNL)\,F(M;...)\nonumber\\
& &  +\,\,C_0(z,\fNL;M_{\rm min})\frac{\bar{\rho}}{M_{\rm min}}\,F(M_{\rm min};...).
\eea
We have found that the predictions from the Halo Model are insensitive to decreasing further the lower bounds below $M_{\rm min} = 10^6 \Ms$, so we have taken this value as the minimum mass cut-off.

Other constraint equations, involving the bias functions, are given by 
\bea\label{eq:condbiasNGsi_v2}
{1\over\bar{\rho}}\int \!\! dm\,m\,n_{\rm NG}\,\left(b_{i,{\rm g}}+\Delta b_{i,{\rm NG}}^{(a)}\right)  &=&  \delta_{i1}\,,\\
{1\over\bar{\rho}}\int \!\! dm\,m\,n_{\rm NG}\,\Delta b_{i,{\rm NG}}^{(b)} &=& 0 \,,~\forall\,i \geqslant1\,,
\label{eq:condbiasNGsd_v2}
\eea
where $\Delta b_{i,{\rm NG}}^{(a)}$ are the scale-independent (\rm SI) corrections to $b_{i}$ and $\Delta b_{i,{\rm NG}}^{(b)}$ the scale-dependent (\rm SD) ones. Again, we enforce these constraints by adding extra terms as
\bea\label{eq:condbiasNGsi_v2}
&& {1\over\bar{\rho}}\int \!\! dm\,m\,n_{\rm NG}\,\left(b_{i,{\rm g}}+\Delta b_{i,{\rm NG}}^{(a)}\right) \nonumber\\ && \hspace{1cm} = {1\over\bar{\rho}}\int_{M_{\rm min}}^{M_{\rm max}}\!\! dm\,m\,n_{\rm NG}\,\left(b_{i,{\rm g}}+\Delta b_{i,{\rm NG}}^{(a)}\right) + C_{i}^{(\rm SI)}(m,z,\fNL;M_{\rm min})\frac{M_{\rm min}}{\bar\rho} = \delta_{i1}\,,\\
&& {1\over\bar{\rho}}\int \!\! dm\,m\,n_{\rm NG}\,\Delta b_{i,{\rm NG}}^{(b)} \nonumber\\
&& \hspace{1cm} = {1\over\bar{\rho}}\int_{M_{\rm min}}^{M_{\rm max}}\!\! dm\,m\,n_{\rm NG}\,\Delta b_{i,{\rm NG}}^{(b)} + C_{i}^{(\rm SD)}(m,z,\fNL,k;M_{\rm min})\frac{M_{\rm min}}{\bar\rho} = 0 \,,~\forall\,i \geqslant1\,,
\label{eq:condbiasNGsd_v2}
\eea
where the new constants read
\bea
C_i^{(\rm SI)}(z,\fNL;M_{\rm min}) = [\delta_{i1}-D_i^{(\rm SI)}(z,\fNL;M_{\rm min})]\frac{\bar{\rho}}{M_{\rm min}}\,,\\
D_i^{(\rm SI)}(z,\fNL;M_{\rm min}) \equiv \frac{1}{\bar{\rho}}\int_{M_{\rm min}}^{M_{\rm max}} \!\! dM\,M\,n_{\rm NG}\,\left(b_{i,{\rm g}}+\Delta b_{i,{\rm NG}}^{(a)}\right)
\eea
and
\bea
C_i^{(\rm SD)}(z,\fNL,k;M_{\rm min}) = - D_i^{(\rm SD)}(z,\fNL,k;M_{\rm min})\frac{\bar{\rho}}{M_{\rm min}}\,,\\
D_i^{(\rm SD)}(z,\fNL,k;M_{\rm min}) \equiv \frac{1}{\bar{\rho}}\int_{M_{\rm min}}^{M_{\rm max}} \!\! dM\,M\,n_{\rm NG}\,\Delta b_{i,{\rm NG}}^{(b)}.
\eea
Integrating over the mass the product of the mass function $n(M)$ with any bias function (or its non-Gaussian corrections) times some other function $F(M)$, we apply an analogous prescription like
\bea\label{eq:IRmassCorrForBiasFunct}
&& \int \!\! dM\,n_{\rm NG}\,\left(b_{i,{\rm g}}+\Delta b_{i,{\rm NG}}^{(a)}\right)\,F(M;...) \nonumber \\ 
&& \hspace*{0.5 cm} = \int_{M_{\rm min}}^{M_{\rm max}} \!\! dM\,M\,n_{\rm NG}\,\left(b_{i,{\rm g}}+\Delta b_{i,{\rm NG}}^{(a)}\right)\,F(M,...) + C_i^{(\rm SI)}(z,\fNL;M_{\rm min})\frac{\bar{\rho}}{M_{\rm min}}\,F(M_{\rm min};...),\\
&&\int \!\! dM\,n_{\rm NG}\,\Delta b_{i,{\rm NG}}^{(b)}\,F(M;...) \nonumber \\
&&\hspace*{0.5 cm} = \int_{M_{\rm min}}^{M_{\rm max}} \!\! dM\,M\,n_{\rm NG}\,\Delta b_{i,{\rm NG}}^{(b)}\,F(M,...) + C_i^{(\rm SD)}(z,\fNL,k;M_{\rm min})\frac{\bar{\rho}}{M_{\rm min}}\,F(M_{\rm min};...).
\eea
We have checked that the predictions from the Halo Model where the addition of these extra constants are involved, are again insensitive to decreasing further the lower bound below $M_{\rm min} = 10^6 \Ms$. Thus we have choosen that value as the minimum mass cut-off for any mass integral all the Halo Model terms.
Finally note that after introducing $C_0$, the additional assumption of $C_{i}{(\rm SI)}$ and $C_{i}{(\rm SD)}$ was necessary and not redundant, since the latter are not directly related to $C_0$. Should any other constraint equation in the form of a mass integral over the mass function or the bias functions times other (mass dependent) functions exist, then another 'compensating' constant should be defined following the same philosophy.

\setlength{\bibsep}{0.5pt}

%\bibliography{Bibliography}

\begin{thebibliography}{91}
\expandafter\ifx\csname natexlab\endcsname\relax\def\natexlab#1{#1}\fi

\bibitem[{Achitouv \& Corasaniti(2012)}]{AchitouvCorasaniti2012}
Achitouv, I., \& Corasaniti, P. 2012, \jcap, 2, 2

\bibitem[{Afshordi \& Tolley(2008)}]{AfshordiTolley2008}
Afshordi, N., \& Tolley, A.~J. 2008, \prd, 78, 123507

\bibitem[{Baldauf {et~al.}(2011)Baldauf, Seljak, \&
  Senatore}]{BaldaufSeljakSenatore2011}
Baldauf, T., Seljak, U., \& Senatore, L. 2011, \jcap, 4, 6

\bibitem[{Bartolo {et~al.}(2010)Bartolo, {Beltr{\'a}n Almeida}, Matarrese,
  Pietroni, \& Riotto}]{BartoloEtal2010}
Bartolo, N., {Beltr{\'a}n Almeida}, J.~P., Matarrese, S., Pietroni, M., \&
  Riotto, A. 2010, \jcap, 3, 11

\bibitem[{Bartolo {et~al.}(2004)Bartolo, Komatsu, Matarrese, \&
  Riotto}]{BartoloEtal2004}
Bartolo, N., Komatsu, E., Matarrese, S., \& Riotto, A. 2004, \physrep, 402, 103

\bibitem[{Bartolo {et~al.}(2002)Bartolo, Matarrese, \&
  Riotto}]{BartoloMatarreseRiotto2002}
Bartolo, N., Matarrese, S., \& Riotto, A. 2002, \prd, 65, 103505

\bibitem[{Bernardeau {et~al.}(2002)Bernardeau, Colombi, Gazta{\~n}aga, \&
  Scoccimarro}]{BernardeauEtal2002}
Bernardeau, F., Colombi, S., Gazta{\~n}aga, E., \& Scoccimarro, R. 2002,
  \physrep, 367, 1

\bibitem[{Bernardeau {et~al.}(2010)Bernardeau, Crocce, \&
  Sefusatti}]{BernardeauCrocceSefusatti2010}
Bernardeau, F., Crocce, M., \& Sefusatti, E. 2010, \prd, 82, 083507

\bibitem[{Bernardeau \& Uzan(2002)}]{BernardeauUzan2002}
Bernardeau, F., \& Uzan, J.-P. 2002, \prd, 66, 103506

\bibitem[{{Bond} {et~al.}(1991){Bond}, {Cole}, {Efstathiou}, \&
  Kaiser}]{BondEtal1991}
{Bond}, J.~R., {Cole}, S., {Efstathiou}, G., \& Kaiser, N. 1991, \apj, 379, 440

\bibitem[{{Bond} {et~al.}(2009){Bond}, {Frolov}, {Huang}, \&
  {Kofman}}]{BondEtal2009}
{Bond}, J.~R., {Frolov}, A.~V., {Huang}, Z., \& {Kofman}, L. 2009, Physical
  Review Letters, 103, 071301

\bibitem[{Bullock {et~al.}(2001)Bullock, Kolatt, Sigad, Somerville, Kravtsov,
  Klypin, Primack, \& Dekel}]{BullockEtal2001}
Bullock, J.~S., Kolatt, T.~S., Sigad, Y., Somerville, R.~S., Kravtsov, A.~V.,
  Klypin, A.~A., Primack, J.~R., \& Dekel, A. 2001, \mnras, 321, 559

\bibitem[{Byrnes \& Choi(2010)}]{ByrnesChoi2010}
Byrnes, C.~T., \& Choi, K.-Y. 2010, {\em Advances in Astronomy}, 2010

\bibitem[{Carbone {et~al.}(2010)Carbone, Mena, \& Verde}]{CarboneMenaVerde2010}
Carbone, C., Mena, O., \& Verde, L. 2010, \jcap, 7, 20

\bibitem[{{Chambers} \& {Rajantie}(2008)}]{ChambersRajantie2008}
Chambers, A., \& {Rajantie}, A. 2008, Journal of Cosmology and Astro-Particle
  Physics, 8, 2

\bibitem[{Chen(2010)}]{Chen2010}
Chen, X. 2010, {\em Advances in Astronomy}, 2010

\bibitem[{Cole \& Kaiser(1989)}]{ColeKaiser1989}
Cole, S., \& Kaiser, N. 1989, \mnras, 237, 1127

\bibitem[{Cooray \& Sheth(2002)}]{CooraySheth2002}
Cooray, A., \& Sheth, R.~K. 2002, \physrep, 372, 1

\bibitem[{Corasaniti \& Achitouv(2011{\natexlab{a}})}]{CorasanitiAchituov2011B}
Corasaniti, P., \& Achitouv, I. 2011{\natexlab{a}}, \prd, 84, 023009

\bibitem[{Corasaniti \& Achitouv(2011{\natexlab{b}})}]{CorasanitiAchitouv2011A}
---. 2011{\natexlab{b}}, Physical Review Letters, 106, 241302

\bibitem[{{Courtin} {et~al.}(2011){Courtin}, {Rasera}, {Alimi}, Corasaniti,
  {Boucher}, \& {F{\"u}zfa}}]{CourtinEtal011}
{Courtin}, J., {Rasera}, Y., {Alimi}, J.-M., Corasaniti, P., {Boucher}, V., \&
  {F{\"u}zfa}, A. 2011, \mnras, 410, 1911

\bibitem[{Crocce {et~al.}(2010)Crocce, Fosalba, Castander, \&
  Gazta{\~n}aga}]{CrocceEtal2010}
Crocce, M., Fosalba, P., Castander, F.~J., \& Gazta{\~n}aga, E. 2010, \mnras,
  403, 1353

\bibitem[{Crocce \& Scoccimarro(2008)}]{CrocceScoccimarro2008}
Crocce, M., \& Scoccimarro, R. 2008, \prd, 77, 023533

\bibitem[{Dalal {et~al.}(2008)Dalal, Dor{\'e}, Huterer, \&
  Shirokov}]{DalalEtal2008}
Dalal, N., Dor{\'e}, O., Huterer, D., \& Shirokov, A. 2008, \prd, 77, 123514

\bibitem[{D'Amico {et~al.}(2011)D'Amico, Musso, Nore{\~n}a, \&
  Paranjape}]{DAmicoEtal2011}
D'Amico, G., Musso, M., Nore{\~n}a, J., \& Paranjape, A. 2011, \jcap, 2, 1

\bibitem[{De~Simone {et~al.}(2011)De~Simone, Maggiore, \&
  Riotto}]{DeSimoneMaggioreRiotto2011}
De~Simone, A., Maggiore, M., \& Riotto, A. 2011, \mnras, 412

\bibitem[{Desjacques {et~al.}(2011{\natexlab{a}})Desjacques, Jeong, \&
  Schmidt}]{DesjacquesJeongSchmidt2011A}
Desjacques, V., Jeong, D., \& Schmidt, F. 2011{\natexlab{a}}, \prd, 84, 061301

\bibitem[{Desjacques {et~al.}(2011{\natexlab{b}})Desjacques, Jeong, \&
  Schmidt}]{DesjacquesJeongSchmidt2011B}
---. 2011{\natexlab{b}}, \prd, 84, 063512

\bibitem[{Desjacques \& Seljak(2010)}]{DesjacquesSeljak2010B}
Desjacques, V., \& Seljak, U. 2010, {\em Advances in Astronomy}, 2010

\bibitem[{Desjacques {et~al.}(2009)Desjacques, Seljak, \&
  Iliev}]{DesjacquesSeljakIliev2009}
Desjacques, V., Seljak, U., \& Iliev, I.~T. 2009, \mnras, 631

\bibitem[{Enqvist \& Sloth(2002)}]{EnqvistSloth2002}
Enqvist, K., \& Sloth, M.~S. 2002, Nuclear Physics B, 626, 395

\bibitem[{Enqvist {et~al.}(2004)Enqvist, Jokinen, \&
  Mazumdar}]{EnqvistEtal2004}
Enqvist, K., Jokinen, A., Mazumdar, A., Multamaki, T., \& Vaihkonen, A. 2004, \prl, 94, 161301

\bibitem[{Fedeli \& Moscardini(2010)}]{FedeliMoscardini2010}
Fedeli, C., \& Moscardini, L. 2010, \mnras, 454

\bibitem[{Figueroa {et~al.}(2010)Figueroa, Caldwell, \&
  Kamionkowski}]{FigueroaCaldwellKamionkowski2010}
Figueroa, D.~G., Caldwell, R.~R., \& Kamionkowski, M. 2010, \prd, 81, 123504

\bibitem[{Fry \& Gazta{\~n}aga(1993)}]{FryGaztanaga1993}
Fry, J.~N., \& Gazta{\~n}aga, E. 1993, \apj, 413, 447

\bibitem[{Giannantonio \& Porciani(2010)}]{GiannantonioPorciani2010}
Giannantonio, T., \& Porciani, C. 2010, \prd, 81, 063530

\bibitem[{Giannantonio {et~al.}(2011)Giannantonio, Porciani, Carron, Amara, \&
  Pillepich}]{GiannantonioEtal2011}
Giannantonio, T., Porciani, C., Carron, J., Amara, A., \& Pillepich, A. 2011,
  arXiv:1109.0958

\bibitem[{Grossi {et~al.}(2007)Grossi, Dolag, Branchini, Matarrese, \&
  Moscardini}]{GrossiEtal2007}
Grossi, M., Dolag, K., Branchini, E., Matarrese, S., \& Moscardini, L. 2007,
  \mnras, 382, 1261

\bibitem[{Grossi {et~al.}(2009)Grossi, Verde, Carbone, Dolag, Branchini,
  Iannuzzi, Matarrese, \& Moscardini}]{GrossiEtal2009}
Grossi, M., Verde, L., Carbone, C., Dolag, K., Branchini, E., Iannuzzi, F.,
  Matarrese, S., \& Moscardini, L. 2009, \mnras, 398, 321

\bibitem[{{Jenkins} {et~al.}(1998){Jenkins}, {Frenk}, {Pearce}, {Thomas},
  {Colberg}, {White}, {Couchman}, {Peacock}, {Efstathiou}, \&
  {Nelson}}]{JenkinsEtal1998}
{Jenkins}, A. {et~al.} 1998, \apj, 499, 20

\bibitem[{Komatsu {et~al.}(2009)Komatsu, Afshordi, Bartolo, Baumann, {Bond},
  {Buchbinder}, Byrnes, Chen, {Chung}, Cooray, Creminelli, Dalal, Dore,
  Easther, {Frolov}, {Gorski}, Jackson, Khoury, {Kinney}, {Kofman}, Koyama,
  {Leblond}, {Lehners}, {Lidsey}, Liguori, Lim, Linde, Lyth, Maldacena,
  Matarrese, McAllister, McDonald, {Mukohyama}, {Ovrut}, Peiris, Riotto,
  {Rodriguez}, {Sasaki}, Scoccimarro, {Seery}, Sefusatti, Seljak, Senatore,
  Shandera, Shellard, Silverstein, Slosar, Smith, {Starobinsky}, Steinhardt,
  {Takahashi}, Tegmark, Tolley, Verde, Wandelt, Wands, {Weinberg}, {Wyman},
  Yadav, \& Zaldarriaga}]{KomatsuEtal2009A}
Komatsu, E. {et~al.} 2009, arXiv:0902.4759

\bibitem[{Komatsu {et~al.}(2011)Komatsu, Smith, {Dunkley}, {Bennett}, {Gold},
  {Hinshaw}, {Jarosik}, {Larson}, {Nolta}, {Page}, {Spergel}, {Halpern},
  {Hill}, {Kogut}, {Limon}, {Meyer}, {Odegard}, {Tucker}, {Weiland}, {Wollack},
  \& {Wright}}]{KomatsuEtal2011}
---. 2011, \apjs, 192, 18

\bibitem[{Komatsu \& Spergel(2001)}]{KomatsuSpergel2001}
Komatsu, E., \& Spergel, D.~N. 2001, \prd, 63, 063002

\bibitem[{Kravtsov {et~al.}(2004)Kravtsov, Berlind, Wechsler, Klypin,
  Gottl{\"o}ber, Allgood, \& Primack}]{KravtsovEtal2004}
Kravtsov, A.~V., Berlind, A.~A., Wechsler, R.~H., Klypin, A.~A., Gottl{\"o}ber,
  S., Allgood, B., \& Primack, J.~R. 2004, \apj, 609, 35

\bibitem[{Kulkarni {et~al.}(2007)Kulkarni, Nichol, Sheth, Seo, Eisenstein, \&
  Gray}]{KulkarniEtal2007}
Kulkarni, G.~V., Nichol, R.~C., Sheth, R.~K., Seo, H.-J., Eisenstein, D.~J., \&
  Gray, A. 2007, \mnras, 378, 1196

\bibitem[{Lam \& Sheth(2009)}]{LamSheth2009B}
Lam, T.~Y., \& Sheth, R.~K. 2009, \mnras, 398, 2143

\bibitem[{Liguori {et~al.}(2010)Liguori, Sefusatti, Fergusson, \&
  Shellard}]{LiguoriEtal2010}
Liguori, M., Sefusatti, E., Fergusson, J.~R., \& Shellard, E. P.~S. 2010,
  {\em Advances in Astronomy}, 2010

\bibitem[{Lo~Verde {et~al.}(2008)Lo~Verde, Miller, Shandera, \&
  Verde}]{LoVerdeEtal2008}
Lo~Verde, M., Miller, A., Shandera, S., \& Verde, L. 2008, Journal of Cosmology
  and Astro-Particle Physics, 4, 14

\bibitem[{LoVerde \& Smith(2011)}]{LoVerdeSmith2011}
LoVerde, M., \& Smith, K.~M. 2011, \jcap, 8, 3

\bibitem[{Lyth {et~al.}(2003)Lyth, Ungarelli, \&
  Wands}]{LythUngarelliWands2003}
Lyth, D.~H., Ungarelli, C., \& Wands, D. 2003, \prd, 67, 023503

\bibitem[{Ma \& Fry(2000)}]{MaFry2000}
Ma, C.-P., \& Fry, J.~N. 2000, \apj, 543, 503

\bibitem[{Maggiore \& Riotto(2010{\natexlab{a}})}]{MaggioreRiotto2010A}
Maggiore, M., \& Riotto, A. 2010{\natexlab{a}}, \apj, 711, 907

\bibitem[{Maggiore \& Riotto(2010{\natexlab{b}})}]{MaggioreRiotto2010B}
---. 2010{\natexlab{b}}, \apj, 717, 515

\bibitem[{Maggiore \& Riotto(2010{\natexlab{c}})}]{MaggioreRiotto2010C}
---. 2010{\natexlab{c}}, \apj, 717, 526

\bibitem[{Mo \& White(1996)}]{MoWhite1996}
Mo, H.~J., \& White, S. D.~M. 1996, \mnras, 282, 347

\bibitem[{Musso \& Paranjape(2012)}]{MussoParanjape2012}
Musso, M., \& Paranjape, A. 2012, \mnras, 420, 369

\bibitem[{Navarro {et~al.}(1997)Navarro, Frenk, \&
  White}]{NavarroFrenkWhite1997}
Navarro, J.~F., Frenk, C.~S., \& White, S. D.~M. 1997, \apj, 490, 493

\bibitem[{Paranjape {et~al.}(2011)Paranjape, Gordon, \&
  Hotchkiss}]{ParanjapeGordonHotchkiss2011}
Paranjape, A., Gordon, C., \& Hotchkiss, S. 2011, \prd, 84, 023517

\bibitem[{Paranjape {et~al.}(2012)Paranjape, Lam, \&
  Sheth}]{ParanjapeLamSheth2012}
Paranjape, A., Lam, T.~Y., \& Sheth, R.~K. 2012, \mnras, 420, 1429

\bibitem[{Peacock \& Smith(2000)}]{PeacockSmith2000}
Peacock, J.~A., \& Smith, R.~E. 2000, \mnras, 318, 1144

\bibitem[{Pillepich {et~al.}(2010)Pillepich, Porciani, \&
  Hahn}]{PillepichPorcianiHahn2010}
Pillepich, A., Porciani, C., \& Hahn, O. 2010, \mnras, 402, 191

\bibitem[{{Press} \& {Schechter}(1974)}]{PressSchechter1974}
{Press}, W.~H., \& {Schechter}, P. 1974, \apj, 187, 425

\bibitem[{Refregier \& Teyssier(2002)}]{RefregierTeyssier2002}
Refregier, A., \& Teyssier, R. 2002, \prd, 66, 043002

\bibitem[{{Regan} \& {Shellard}(2010)}]{ReganShellard2010}
{Regan}, D.~M., \& {Shellard}, E.~P.~S. 2010, \prd, 66, 063527

\bibitem[{Salopek \& Bond(1990)}]{SalopekBond1990}
Salopek, D.~S., \& Bond, J.~R. 1990, \prd, 42, 3936

\bibitem[{{Scherrer} \& {Bertschinger}(1991)}]{ScherrerBertschinger1991}
{Scherrer}, R.~J., \& {Bertschinger}, E. 1991, \apj, 381, 349

\bibitem[{Scoccimarro(1997)}]{Scoccimarro1997}
Scoccimarro, R. 1997, \apj, 487, 1

\bibitem[{Scoccimarro(2000)}]{Scoccimarro2000A}
---. 2000, \apj, 542, 1

\bibitem[{Scoccimarro {et~al.}(2012)Scoccimarro, Manera, Hui, \&
  Chan}]{ScoccimarroEtal2012}
Scoccimarro, R., Manera, M., Hui, L., \& Chan, K.~C. 2012, \prd, 85, 083002

\bibitem[{Scoccimarro {et~al.}(2004)Scoccimarro, Sefusatti, \&
  Zaldarriaga}]{ScoccimarroSefusattiZaldarriaga2004}
Scoccimarro, R., Sefusatti, E., \& Zaldarriaga, M. 2004, \prd, 69, 103513

\bibitem[{Scoccimarro {et~al.}(2001)Scoccimarro, Sheth, Hui, \&
  Jain}]{ScoccimarroEtal2001A}
Scoccimarro, R., Sheth, R.~K., Hui, L., \& Jain, B. 2001, \apj, 546, 20

\bibitem[{Sefusatti(2009)}]{Sefusatti2009}
Sefusatti, E. 2009, \prd, 80, 123002

\bibitem[{Sefusatti {et~al.}(2010)Sefusatti, Crocce, \&
  Desjacques}]{SefusattiCrocceDesjacques2010}
Sefusatti, E., Crocce, M., \& Desjacques, V. 2010, \mnras, 721

\bibitem[{Sefusatti {et~al.}(2011)Sefusatti, Crocce, \&
  Desjacques}]{SefusattiCrocceDesjacques2011}
---. 2011, arXiv:1111.6966

\bibitem[{Sefusatti {et~al.}(2012)Sefusatti, Fergusson, Chen, \&
  Shellard}]{SefusattiEtal2012}
Sefusatti, E., Fergusson, J.~R., Chen, X., \& Shellard, E. P.~S. 2012, 
arXiv:1204.6318 [astro-ph.CO]
%in preparation

\bibitem[{Sefusatti \& Komatsu(2007)}]{SefusattiKomatsu2007}
Sefusatti, E., \& Komatsu, E. 2007, \prd, 76, 083004

\bibitem[{Sefusatti {et~al.}(2007)Sefusatti, Vale, Kadota, \&
  Frieman}]{SefusattiEtal2007}
Sefusatti, E., Vale, C., Kadota, K., \& Frieman, J. 2007, \apj, 658, 669

\bibitem[{Seljak(2000)}]{Seljak2000}
Seljak, U. 2000, \mnras, 318, 203

\bibitem[{Sheth \& Tormen(1999)}]{ShethTormen1999}
Sheth, R.~K., \& Tormen, G. 1999, \mnras, 308, 119

\bibitem[{Slosar {et~al.}(2008)Slosar, Hirata, Seljak, Ho, \&
  Padmanabhan}]{SlosarEtal2008}
Slosar, A., Hirata, C., Seljak, U., Ho, S., \& Padmanabhan, N. 2008, Journal of
  Cosmology and Astro-Particle Physics, 8, 31

\bibitem[{Smith {et~al.}(2011)Smith, Desjacques, \&
  Marian}]{SmithDesjacquesMarian2011}
Smith, R.~E., Desjacques, V., \& Marian, L. 2011, \prd, 83, 043526

\bibitem[{Smith {et~al.}(2003)Smith, {Peacock}, {Jenkins}, White, {Frenk},
  {Pearce}, {Thomas}, {Efstathiou}, \& {Couchman}}]{SmithEtal2003}
Smith, R.~E. {et~al.} 2003, \mnras, 341, 1311

\bibitem[{Smith {et~al.}(2008)Smith, Sheth, \&
  Scoccimarro}]{SmithShethScoccimarro2008}
Smith, R.~E., Sheth, R.~K., \& Scoccimarro, R. 2008, \prd, 78, 023523

\bibitem[{Taruya {et~al.}(2008)Taruya, Koyama, \&
  Matsubara}]{TaruyaKoyamaMatsubara2008}
Taruya, A., Koyama, K., \& Matsubara, T. 2008, \prd, 78, 123534

\bibitem[{{The Planck Collaboration}(2006)}]{PLANCK2006}
{The Planck Collaboration}. 2006, arXiv:astro-ph/0604069

\bibitem[{Valageas(2010)}]{Valageas2010}
Valageas, P. 2010, \aap, 514, A46

\bibitem[{Valageas \& Nishimichi(2011)}]{ValageasNishimichi2011B}
Valageas, P., \& Nishimichi, T. 2011, \aap, 532, A4

\bibitem[{Vernizzi \& Wands(2006)}]{VernizziWands2006}
Vernizzi, F., \& Wands, D. 2006, Journal of Cosmology and Astro-Particle
  Physics, 5, 19

\bibitem[{Wagner \& Verde(2012)}]{WagnerVerde2012}
Wagner, C., \& Verde, L. 2012, \jcap, 3, 2

\bibitem[{Wagner {et~al.}(2010)Wagner, Verde, \&
  Boubekeur}]{WagnerVerdeBoubeker2010}
Wagner, C., Verde, L., \& Boubekeur, L. 2010, \jcap, 10, 22

\bibitem[{{Warren} {et~al.}(2006){Warren}, Abazajian, {Holz}, \&
  {Teodoro}}]{WarrenEtal2006}
{Warren}, M.~S., Abazajian, K., {Holz}, D.~E., \& {Teodoro}, L. 2006, \apj,
  646, 881

\bibitem[{Xia {et~al.}(2011)Xia, Baccigalupi, Matarrese, Verde, \&
  Viel}]{XiaEtal2011}
Xia, J.-Q., Baccigalupi, C., Matarrese, S., Verde, L., \& Viel, M. 2011, \jcap,
  8, 33

\end{thebibliography}

\end{document}